\documentclass[twocolumn, times]{aastex63}
\usepackage{multirow}
\usepackage{xcolor}

\shorttitle{MAPS XVIII}
\shortauthors{Teague et al.}

\begin{document}

\title{Molecules with ALMA at Planet-forming Scales (MAPS XVIII):\\Kinematic Substructures in the Disks of HD~163296 and MWC~480}

\author[0000-0002-0786-7307]{Richard Teague}
\affiliation{Center for Astrophysics \textbar{} Harvard \& Smithsonian, 60 Garden Street, Cambridge, MA 02138, USA}

\author[0000-0001-7258-770X]{Jaehan Bae}
\altaffiliation{NASA Hubble Fellowship Program Sagan Fellow}
\affil{Earth and Planets Laboratory, Carnegie Institution for Science, 5241 Broad Branch Road NW, Washington, DC 20015, USA}
\affiliation{Department of Astronomy, University of Florida, Gainesville, FL 32611, USA}

\author[0000-0003-3283-6884]{Yuri Aikawa}
\affil{Department of Astronomy, Graduate School of Science, The University of Tokyo, 7-3-1 Hongo, Bunkyo-ku, Tokyo 113-0033, Japan}

\author[0000-0003-2253-2270]{Sean M. Andrews}
\affiliation{Center for Astrophysics \textbar{} Harvard \& Smithsonian, 60 Garden Street, Cambridge, MA 02138, USA}

\author[0000-0003-4179-6394]{Edwin A.\ Bergin}
\affiliation{Department of Astronomy, University of Michigan, 323 West Hall, 1085 S. University Avenue, Ann Arbor, MI 48109, USA}

\author[0000-0002-8716-0482]{Jennifer B. Bergner} 
\altaffiliation{NASA Hubble Fellowship Program Sagan Fellow}
\affiliation{University of Chicago Department of the Geophysical Sciences, Chicago, IL 60637, USA}

\author[0000-0002-8692-8744]{Yann Boehler}
\affiliation{Univ. Grenoble Alpes, CNRS, IPAG, F-38000 Grenoble, France}

\author[0000-0003-2014-2121]{Alice S. Booth}
\affiliation{Leiden Observatory, Leiden University, 2300 RA Leiden, the Netherlands}
\affiliation{School of Physics and Astronomy, University of Leeds, Leeds LS2 9JT, UK}

\author[0000-0003-4001-3589]{Arthur D. Bosman}
\affiliation{Department of Astronomy, University of Michigan, 323 West Hall, 1085 S. University Avenue, Ann Arbor, MI 48109, USA}

\author[0000-0002-2700-9676]{Gianni Cataldi}
\affil{Department of Astronomy, Graduate School of Science, The University of Tokyo, 7-3-1 Hongo, Bunkyo-ku, Tokyo 113-0033, Japan}
\affil{National Astronomical Observatory of Japan, Osawa 2-21-1, Mitaka, Tokyo 181-8588, Japan}

\author[0000-0002-1483-8811]{Ian Czekala}
\altaffiliation{NASA Hubble Fellowship Program Sagan Fellow}
\affiliation{Department of Astronomy and Astrophysics, 525 Davey Laboratory, The Pennsylvania State University, University Park, PA 16802, USA}
\affiliation{Center for Exoplanets and Habitable Worlds, 525 Davey Laboratory, The Pennsylvania State University, University Park, PA 16802, USA}
\affiliation{Center for Astrostatistics, 525 Davey Laboratory, The Pennsylvania State University, University Park, PA 16802, USA}
\affiliation{Institute for Computational \& Data Sciences, The Pennsylvania State University, University Park, PA 16802, USA}
\affiliation{Department of Astronomy, 501 Campbell Hall, University of California, Berkeley, CA 94720-3411, USA}

\author[0000-0003-4784-3040]{Viviana V. Guzm\'{a}n}
\affiliation{Instituto de Astrof\'isica, Pontificia Universidad Cat\'olica de Chile, Av. Vicu\~na Mackenna 4860, 7820436 Macul, Santiago, Chile}

\author[0000-0001-6947-6072]{Jane Huang}
\altaffiliation{NASA Hubble Fellowship Program Sagan Fellow}
\affiliation{Center for Astrophysics \textbar{} Harvard \& Smithsonian, 60 Garden Street, Cambridge, MA 02138, USA}
\affiliation{Department of Astronomy, University of Michigan, 323 West Hall, 1085 S. University Avenue, Ann Arbor, MI 48109, USA}

\author[0000-0003-1008-1142]{John D. Ilee}
\affiliation{School of Physics and Astronomy, University of Leeds, Leeds LS2 9JT, UK}

\author[0000-0003-1413-1776]{Charles J. Law}
\affiliation{Center for Astrophysics \textbar{} Harvard \& Smithsonian, 60 Garden Street, Cambridge, MA 02138, USA}

\author[0000-0003-1837-3772]{Romane Le Gal}
\affiliation{Center for Astrophysics \textbar{} Harvard \& Smithsonian, 60 Garden Street, Cambridge, MA 02138, USA}
\affiliation{Univ. Grenoble Alpes, CNRS, IPAG, F-38000 Grenoble, France}
\affiliation{IRAP, Universit\'{e} de Toulouse, CNRS, CNES, UT3, 31400 Toulouse, France}
\affiliation{IRAM, 300 rue de la piscine, F-38406 Saint-Martin d'H\`{e}res, France}

\author[0000-0002-7607-719X]{Feng Long}
\affiliation{Center for Astrophysics \textbar{} Harvard \& Smithsonian, 60 Garden Street, Cambridge, MA 02138, USA}

\author[0000-0002-8932-1219]{Ryan A. Loomis}
\affiliation{National Radio Astronomy Observatory, 520 Edgemont Rd., Charlottesville, VA 22903, USA}

\author[0000-0002-1637-7393]{Fran\c cois M\'enard}
\affiliation{Univ. Grenoble Alpes, CNRS, IPAG, F-38000 Grenoble, France}

\author[0000-0001-8798-1347]{Karin I. \"Oberg}
\affiliation{Center for Astrophysics \textbar{} Harvard \& Smithsonian, 60 Garden Street, Cambridge, MA 02138, USA}

\author[0000-0002-1199-9564]{Laura M. P\'erez} \affiliation{Departamento de Astronom\'ia, Universidad de Chile, Camino El Observatorio 1515, Las Condes, Santiago, Chile}

\author[0000-0002-6429-9457]{Kamber R. Schwarz}
\altaffiliation{NASA Hubble Fellowship Program Sagan Fellow}
\affiliation{Lunar and Planetary Laboratory, University of Arizona, 1629 E. University Blvd, Tucson, AZ 85721, USA}

\author[0000-0002-5991-8073]{Anibal Sierra}
\affiliation{Departamento de Astronom\'ia, Universidad de Chile, Camino El Observatorio 1515, Las Condes, Santiago, Chile}

\author[0000-0001-6078-786X]{Catherine Walsh}
\affiliation{School of Physics and Astronomy, University of Leeds, Leeds LS2 9JT, UK}

\author[0000-0003-1526-7587]{David J. Wilner}
\affiliation{Center for Astrophysics \textbar{} Harvard \& Smithsonian, 60 Garden Street, Cambridge, MA 02138, USA}

\author[0000-0003-4099-6941]{Yoshihide Yamato}
\affil{Department of Astronomy, Graduate School of Science, The University of Tokyo, 7-3-1 Hongo, Bunkyo-ku, Tokyo 113-0033, Japan}

\author[0000-0002-0661-7517]{Ke Zhang}
\altaffiliation{NASA Hubble Fellow}
\affiliation{Department of Astronomy, University of Michigan, 323 West Hall, 1085 S. University Avenue, Ann Arbor, MI 48109, USA}
\affiliation{Department of Astronomy, University of Wisconsin-Madison, 475 N Charter St, Madison, WI 53706}

\begin{abstract}
We explore the dynamical structure of the protoplanetary disks surrounding HD~163296 and MWC~480 as part of the Molecules with ALMA at Planet Forming Scales (MAPS) large program. Using the $J = 2-1$ transitions of $^{12}$CO, $^{13}$CO and C$^{18}$O imaged at spatial resolutions of $\sim 0\farcs15$ and with a channel spacing of $200~{\rm m\,s^{-1}}$, we find perturbations from Keplerian rotation in the projected velocity fields of both disks ($\lesssim\!5\%$ of the local Keplerian velocity), suggestive of large-scale (10s of au in size), coherent flows. By accounting for the azimuthal dependence on the projection of the velocity field, the velocity fields were decomposed into azimuthally averaged orthogonal components, $v_{\phi}$, $v_r$ and $v_z$. Using the optically thick $^{12}$CO emission as a probe of the gas temperature, local variations of $\approx\! 3$~K ($\approx\! 5 \%$ relative changes) were observed and found to be associated with the kinematic substructures. The MWC~480 disk hosts a suite of tightly wound spiral arms. The spirals arms, in conjunction with the highly localized perturbations in the gas velocity structure (kinematic planetary signatures), indicate a giant planet, $\sim\! 1 M_{\rm Jup}$, at a radius of $\approx 245$~au. In the disk of HD~163296, the kinematic substructures were consistent with previous studies of \citet{Pinte_ea_2018b} and \citet{Teague_ea_2018a} advocating for multiple $\sim\! 1 M_{\rm Jup}$ planets embedded in the disk. These results demonstrate that molecular line observations that characterize the dynamical structure of disks can be used to search for the signatures of embedded planets. This paper is part of the MAPS special issue of the Astrophysical Journal Supplement.
\end{abstract}


\section{Introduction}
\label{sec:introduction}

It is clear that the Atacama Large Millimeter/submillimeter Array (ALMA) has driven a revolution in our understanding of the physical and chemical structure of protoplanetary disks, the birthplaces of planets. Not only has it revealed a stunning variety and ubiquity of substructures in the mm continuum, such as gaps, rings and vortices \citep[e.g.,][]{ALMA_ea_2015, Andrews_ea_2018, Long_ea_2018}, but also a similar level of complexity in the morphology of molecular line emission \citep[e.g.,][]{Christiaens_ea_2014, Teague_ea_2017, Teague_ea_2019a, Huang_ea_2020}. While the former is likely due to the grains being shepherded by changes in the global gas physical structure \citep{Dullemond_ea_2018}, identifying the cause of changes in molecular emission requires distinguishing between physical and chemical effects \citep{Facchini_ea_2018, vanderMarel_ea_2018}.

The unparalleled sensitivity of ALMA has enabled an alternative approach to detailing the physical structure of the gas: the gas dynamics. A significant advantage of a dynamical study of the planet formation environment is that it circumvents the need to account for the chemical effects when relating measured fluxes to column densities of H$_2$ gas. This is most clearly seen with the pressure-modulation of the rotation velocity,
\begin{equation}
    \frac{v_{\phi}^2}{r} = \frac{GM_{*} r}{(r^2 + z^2)^{3/2}} + \frac{1}{\rho_{\rm gas}} \frac{\partial P}{\partial r},
    \label{eq:v_phi}
\end{equation}
\noindent where $r$ and $z$ are the cylindrical radius and vertical height above the midplane, respectively, $M_{*}$ is the stellar mass, $\rho_{\rm gas}$ is the gas density, dominated by H$_2$, and $P$ is the gas pressure. Changes in the local gas pressure, or gas surface density, will therefore manifest as changes in $v_{\phi}$, irrespective of the abundance of the observed molecule. The advantage of this approach is clear: variations in the total gas column can be inferred without the need for expensive astrochemical models. For example, \citet{Teague_ea_2018a, Teague_ea_2018c} detected subtle variations in $v_{\phi}$ for the disks around HD~163296 and AS~209 which was used to model the perturbed gas surface density profile.

It is well understood that giant planets will open up gaps in the gas distribution of their parental protoplanetary disks when they reach a sufficient mass \citep[e.g.,][]{Papaloizou_Lin_1984}. While this is an attractive, and frequently invoked, scenario to explain perturbations in the gas surface density, other dynamical processes, such as the magneto-rotational instability \citep[MRI;][]{Flock_ea_2015} or the vertical shear instability \citep[VSI;][]{Flock_ea_2017}, have been demonstrated to result in comparable perturbations. A way to distinguish between these scenarios is leverage the fact that planets should only drive \emph{localized} perturbations, as the shocks from their spirals are strongest in their immediate vicinity. Such localized perturbations ($\sim\!10\%$ of the background Keplerian rotation) in the velocity structure of HD~163296 and HD~97048 were reported by \citet{Pinte_ea_2018b, Pinte_ea_2019}, which the authors showed were consistent with giant, ${\sim}~2~{\rm M_{Jup}}$ planets.

Embedded planets are also expected to drive large velocity perturbations in all three components of the gas velocity, $v_{\phi}$, $v_r$ and $v_z$, \citep[e.g.,][]{Perez_ea_2015, Perez_ea_2018, Pinte_ea_2019, DiskDynamics_ea_2020}. These can be probed by decomposing the line-of-sight velocity $v_0$, which is the sum of the systemic velocity, $v_{\rm LSR}$, and the projected disk velocity field, $v_{\rm disk}$, into the three cardinal velocity components,
\begin{equation}
    v_{\rm disk} = v_{\phi} \cos(\phi) \sin(i) + v_{\rm r} \sin(\phi) \sin(i) + v_{\rm z} \cos(i)
    \label{eq:v_disk}
\end{equation}
\noindent where $\phi$ is the azimuthal angle of the disk (measured such that $\phi = 0$ is the red-shifted major axis) and $i$ is the inclination of the disk. \citet{Teague_ea_2019b, Teague_ea_2019a} demonstrated that these components can be disentangled if we know their geometric projection and assume azimuthal asymmetry. A similar approach is frequently used in studies of galactic dynamics \citep[e.g.,][]{Krajnovic_ea_2006}.

There has been much recent success in using the high spatial and spectral observations from ALMA to probe the gas dynamics and infer the presence of unseen planets by the detection of kinematic planetary signatures \citep[KPS;][]{DiskDynamics_ea_2020}. This term encompasses all localized features in the gas dynamical structure which are driven by an unseen planetary perturber, such as `kinks' or `wiggles' in the emission morphology or `Doppler-flips' and spiral patterns in rotation map residuals. \citet{Pinte_ea_2018b} found localized velocity deviations in the outer disk of HD~163296, consistent with a $\sim\!2~M_{\rm Jup}$ planet at a radius of $\approx 260$~au. In the same source, \citet{Teague_ea_2019b} used $^{12}$CO emission to measure the azimuthally averaged 3D velocity structure, revealing a large gas pressure minimum at the orbital radius of the planet and evidence for meridional flows; the motion of gas from the atmosphere of the disk into (likely planet opened) gaps \citep{Morbidelli_ea_2014, Szulagyi_ea_2014}. Other KPS have been found in AB~Aur \citep{Tang_ea_2017}, HD~97048 \citep{Pinte_ea_2019}, HD~100546 \citep{Casassus_Perez_2019, Perez_ea_2020}, TW~Hya \citep{Teague_ea_2019a}, along with several tentative features in sources targeted by the DSHARP survey \citep{Pinte_ea_2020}. These results demonstrate that we are not only on the verge of painting a six dimensional view of the planet formation process, but that we have the potential to transform ALMA into a \mbox{(sub-)}mm planet-hunting instrument. Observations of molecular line emission will enable for a unique quantification of the planet-disk interactions in a less ambiguous way than with \mbox{(sub-)mm} continuum alone.

The Molecules with ALMA at Planet-forming Scales (MAPS) large program \citep{Oberg_ea_2020} produced some of the deepest and highest angular resolution observations of molecular line emission from protoplanetary disks to date with a view to characterize the chemical complexities associated with the planet formation process. These observations are ideal for also characterizing the dynamical structure of the planet-hosting disks. In this paper, we use these data to analyze the velocity structures of the disks around HD~163296 and MWC~480, the two Herbig Ae stars in the program. These two sources were selected as the other three MAPS sources contained complex $^{12}$CO emission morphologies: GM~Aur was found to have highly complex extended emission, presented in \citet[MAPS~XIX]{Huang_MAPS}; AS~209 has substantial cloud contamination over a large regions of the disk \citep[e.g.,][]{Huang_ea_2016} and IM~Lup has extensive diffuse $^{12}$CO emission, potentially associated with a photoevaporative wind \citep[e.g.,][]{Cleeves_ea_2016}. In Section~\ref{sec:observations} we describe the observations used, and measure the line-of-sight velocities for the bright $^{12}$CO emission in Section~\ref{sec:rotation_maps}. In Section~\ref{sec:velocity_profiles} these are disentangled into azimuthally symmetric orthogonal components. Section~\ref{sec:gas_temperature} describes the local changes in the gas temperature which are associated with the perturbations in the gas dynamics. Finally, these perturbations are discussed in the context of embedded planets in Section~\ref{sec:discussion}, with the results summarized in Section~\ref{sec:summary}.

\section{Observations}
\label{sec:observations}

\begin{figure*}
    \centering
    \includegraphics[width=0.8\textwidth]{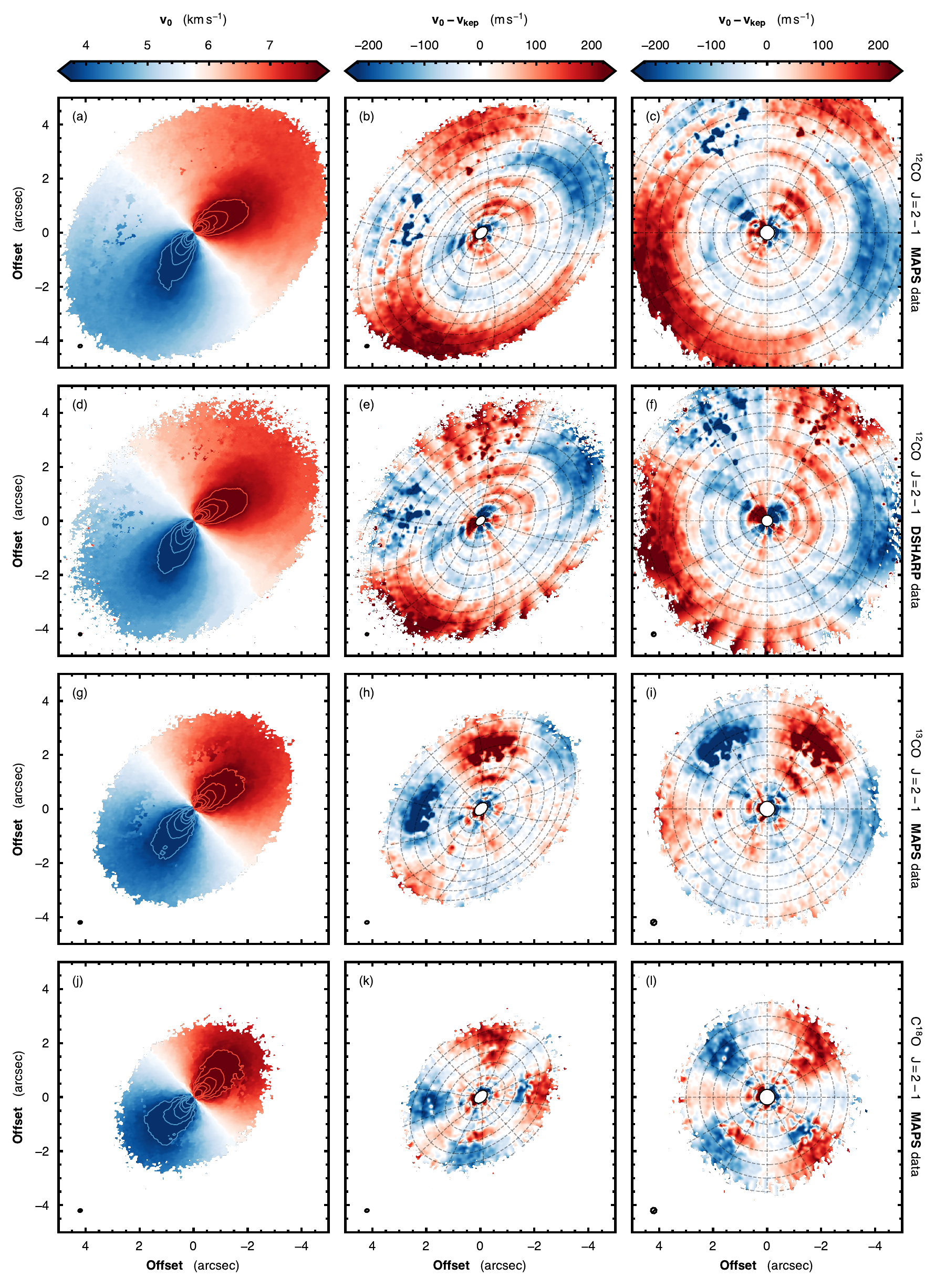}
    \caption{Gallery of $v_0$ maps for different $J = 2-1$ CO isotopologue emission from the disk around HD~163296 using both MAPS and DSHARP data \citep{Andrews_ea_2018, Isella_ea_2018}. The left column shows the raw $v_0$ map, the central column shows the residuals after subtracting a best-fit Keplerian rotation model, and the right column shows the residuals deprojected into a face-on orientation accounting for the elevated emission surface. Solid colored contours in the left panel show iso-velocity contours of $v_{\rm LSR} \pm \{2,\, 2.5,\, 3,\,...\}~{\rm km\,s^{-1}}$. Dashed black contours in the central and right hand columns show contours of constant radius and azimuth in the disk frame, with intervals of $0\farcs5$ and 30\degr{}, respectively. The beam size is shown in the bottom left of each panel. For the deprojected maps in the right hand column, the beam size is only representative as the deprojection results in a non-uniform spatial resolution across the image. The disk rotates clockwise in all panels.
    }
    \label{fig:HD163296_v0maps}
\end{figure*}

As part of the MAPS program, both HD~163296 and MWC~480 were observed with the correlator tuned to cover the $J = 2-1$ rotational transitions of $^{12}$CO, $^{13}$CO and C$^{18}$O as to trace a range of vertical heights in the disk \citep[MAPS~IV]{Law_ea_2020_surf}. We refer the reader to \citet[MAPS~I]{Oberg_ea_2020} for a full description of the observational setup and calibration process and \citet[MAPS~II]{Czekala_ea_2020} for a thorough description of the imaging process.

For this paper we use the \texttt{robust=0.5} weighted, JvM-corrected\footnote{A correction related to the mismatch in \texttt{CLEAN} beam and dirty beam sizes when combining a \texttt{CLEAN} image and the residuals \citep[MAPS~II]{Jorsater_vanMoorsel_1995, Czekala_ea_2020}.} images as these produced the smallest synthesized beam. This resulted in beam sizes of approximately $0\farcs14 \times 0\farcs11$ at a position angle of 104\degr{} for the CO isotopologue emission in HD~163296 and $0\farcs17 \times 0\farcs12$ at a position angle of 6\degr{} for isotopologue emission in MWC~480. All images were produced with a channel spacing of $200~{\rm m\,s^{-1}}$. The RMS was measured as the standard deviation of pixel values in a circular area of 1\arcsec{} in radius centered on the phase center of the first and last channel ($\pm 12~{\rm km\,s^{-1}}$ from the systemic velocity) in the image cube. For HD~163296, the RMS was measured to be 0.58, 0.54 and $0.37~{\rm mJy~beam^{-1}}$ for $^{12}$CO, $^{13}$CO and C$^{18}$O, respectively. In MWC~480, these were 0.73, 0.68 and $0.36~{\rm mJy~beam^{-1}}$. A set of channel maps for the $^{12}$CO, $^{13}$CO and C$^{18}$O emission can be found in Appendix~\ref{sec:app:channel_maps}.

\section{Rotation Maps}
\label{sec:rotation_maps}

In this section, we collapse the image cubes along the spectral axis to study their projected velocity fields and search for any deviations relative to a background Keplerian model.

\subsection{Method}
\label{sec:rotation_maps:method}

We use the `quadratic' method implemented in \texttt{bettermoments} to produce maps of the line center, $v_0$ maps, including a statistical uncertainty for the measured $v_0$ as described in \citet{Teague_Foreman-Mackey_2018}. The maps are then masked to only include regions where the peak intensities are greater than 5 times the RMS value measured in a line free channel in order to remove noisy values at the outer edge of the disk. We explored whether an additional smoothing along the spectral axis prior to the collapse of the data cube would improve the fidelity of the rotation map (as advocated by \citealt{Teague_Foreman-Mackey_2018} for low signal-to-noise ratio spectra), however we found that this did not significantly improve the resulting $v_0$ maps. Appendix~\ref{sec:app:creation_of_rotation_maps} discusses this in more detail.

\begin{figure}
    \centering
    \includegraphics[width=\columnwidth]{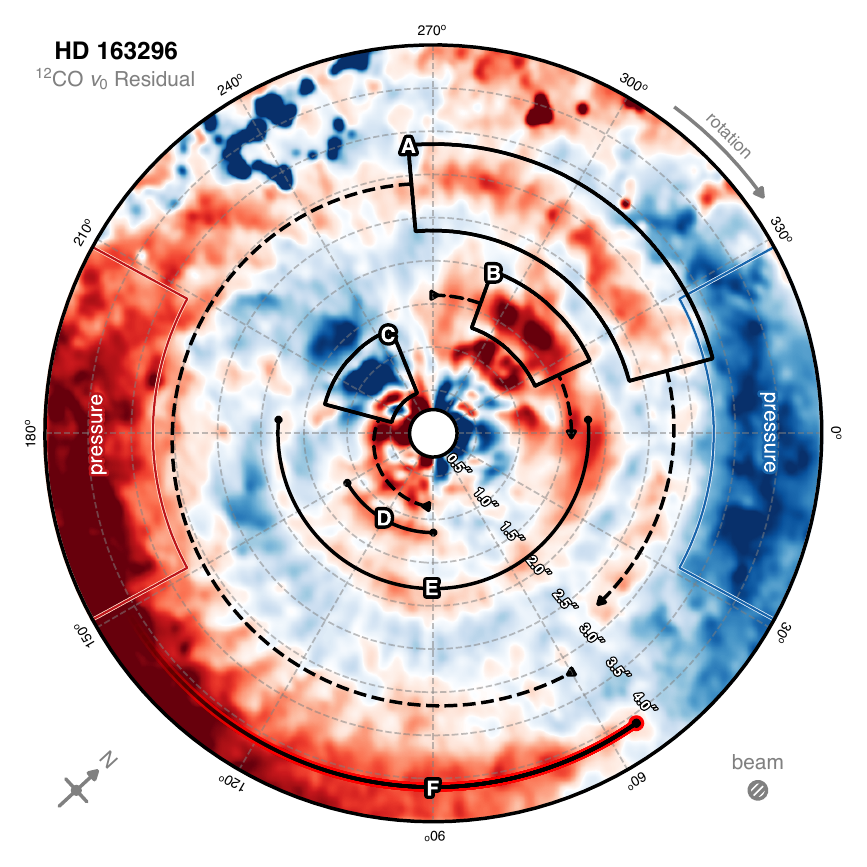}
    \caption{A zoom-in of Fig.~\ref{fig:HD163296_v0maps}c. A representative beam size is shown in the lower right, however due to the elevated emission surface, the true angular resolution varies across the map. The figure extends out to $4\farcs4$, and the color scaling ranges between residuals of $\pm 180~{\rm m\,s^{-1}}$. Residuals due to the pressure gradient slowing the rotation of the disk are observed along the major axis, marked as ``pressure''. Three regions of large velocity deviations are marked by black boxes, centered at $(r,\, \phi) = (2\farcs85,\, 305\degr)$, A, $(1\farcs65,\, 315\degr)$, B, and $(0\farcs9,\, 220\degr)$, C, all of which have extended structure shown by the dashed lines. The outer feature, A, is associated with the planet identified by \citet{Pinte_ea_2018b}, while the in two, B and C, are likely associated with the potential ``kink'' reported by \citet{Pinte_ea_2020}, or due to the asymmetry in the continuum emission \citep{Isella_ea_2018}. Two arc-like structures are also seen at radii of $1\farcs15$ (116~au), and $1\farcs8$ (182~au), shown by the solid lines, labeled D and E, respectively, in addition to a larger arc feature at $4\farcs1$ (414~au), shown by the thick solid line with a faded end, annotated as F. It is unclear where this outer-most arc ends as the perturbation extends into a a region dominated by pressure-modulated deviations.}
    \label{fig:HD163296_v0maps_annotated}
\end{figure}

To each of these maps we fit a Keplerian rotation model, $v_{\rm kep}$, given by Eqn.~\ref{eq:v_phi} when assuming $\partial P / \partial r = 0$. To find the best-fit Keplerian velocity field, $v_{\rm kep}$, we use the Python package \texttt{eddy} \citep{eddy} to explore the posterior distributions of the model parameters. For these models there is a strong degeneracy between the disk inclination and the stellar mass that can usually be broken with high resolution observations such as those from the MAPS project. However, if there are perturbations in the velocity structure, these will dominate the residuals and thus preclude a tight constraint on the inclination. As such, incorporating prior information on the disk inclination, from the continuum emission, for example, is preferable. We adopt the inclination and binned $z(r)$ profiles from \citet{Huang_ea_2018a} and \citet[MAPS~IV]{Law_ea_2020_surf}, respectively, for HD~163296. For MWC~480, we found that the velocity structure in the outer disk is more consistent with an inclination of 33\degr{}, rather than the 37\degr{} based on continuum fitting \citep{Liu_ea_2019} and that was adopted for other MAPS papers. As systematic residuals from a misspecified inclination dominate our residual map \citep[see][for example]{Yen_Gu_2020}, we choose to use the 33\degr{} inclination for the entire MWC~480 disk and recalculate the $z(r)$ profiles following the method described in \citet[MAPS~IV]{Law_ea_2020_surf}.  Appendix~\ref{sec:app:MWC480_inclination} provides more details on this. Other geometrical properties, including source center, $(x_0,\,y_0)$, position angle, PA, dynamical mass, $M_*$, and systemic velocity, $v_{\rm LSR}$, are left free. Appendix~\ref{sec:app:rotation_map_fitting} discusses the model fitting procedure in more detail.

\subsection{HD~163296}
\label{sec:rotation_maps:HD163296}

The $v_0$ maps of HD~163296 for all three CO isotopologues are shown in the left hand column of Figure~\ref{fig:HD163296_v0maps}, with each isotopologue on a separate row. The color scale has been saturated in the central regions to bring out the detail along the minor axes of the disk. Solid contour lines show iso-velocity contours in steps of $500~{\rm m\,s^{-1}}$ in this region. The central column shows the residual after subtracting a best-fit Keplerian mode, $v_{\rm kep}$. To suppress the noise in the residual map, these are smoothed by convolution with a circular Gaussian kernel with a FWHM of 2 pixels (see Appendix~\ref{sec:app:creation_of_rotation_maps} for a comparison of how the smoothing affects the residuals). Overlaid are contours of constant radius and azimuth in the disk frame of reference, in steps of $0\farcs5$ and $30\degr$, respectively. In the inner $\approx 0\farcs3$ ($\approx 30$~au), beam convolution effects preclude a good fit of the data and are therefore masked in the residual maps. In the right hand column, the residual map has been deprojected into a face-on view, taking account of the disk inclination, position angle and emission surface. The north-facing minor axis is orientated along the positive $y$-axis. In all these panels, the disk is rotating in a clockwise fashion. Note that for HD~163296, continuum emission is only detected out to a radius of ${\sim} 1\farcs7$ \citep{Isella_ea_2016}, such that the gas is over 3 times larger in radius.

We compare the rotation maps for $^{12}$CO emission made with MAPS data to those made with data from DSHARP \citep{Andrews_ea_2018, Isella_ea_2018} in Fig.~\ref{fig:HD163296_v0maps} for HD~163296. The DSHARP data was taken at a slightly higher angular resolution ($\approx\! 100$~mas versus $\approx\! 125$~mas for MAPS), but at a lower effective velocity resolution ($640~{\rm m\,s^{-1}}$ versus the binned down $200~{\rm m\,s^{-1}}$ of the MAPS data). There are two striking features in this comparison. First, is the extraordinary similarity in the observed residuals, despite being entirely independent data sets. This strongly suggests that with ALMA we are hitting the `confusion limit' for rotation maps: that is, our residuals are now dominated by \emph{real} variations in the gas kinematics rather than thermal noise. Secondly, there is considerable ``feathering'' (radial features which result in a beating pattern as a function of azimuth) observed in the DSHARP data that is due to the channelization of the data. This is exacerbated by the low effective velocity resolution of the data ($640~{\rm m\,s^{-1}}$) and is mostly removed with the finer effective velocity resolution achieved by MAPS \citep[MAPS~I]{Oberg_ea_2020}.

\begin{figure*}
    \centering
    \includegraphics[width=0.85\textwidth]{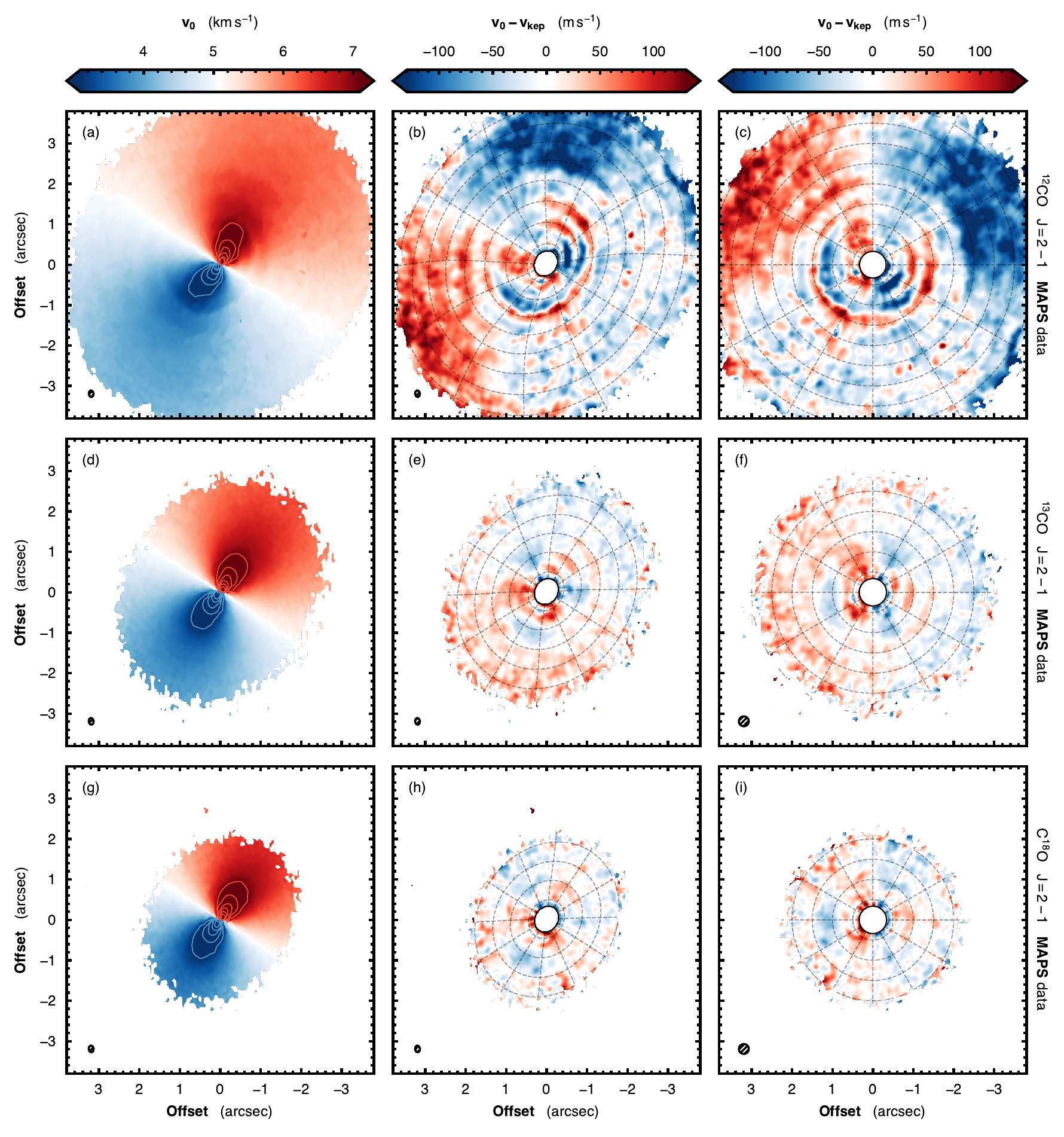}
    \caption{As Fig.~\ref{fig:HD163296_v0maps}, but for MWC~480. As this disk was not targeted by the DSHARP survey, only the MAPS data are shown. The disk rotates in an anti-clockwise direction in all frames. The beam size is shown in the bottom left of each panel. For the deprojected maps in the right hand column, the beam size is only representative as the deprojection results in a non-uniform spatial resolution across the image.}
    \label{fig:MWC480_v0maps}
\end{figure*}

We can identify several dominant features in the residual map of $^{12}$CO emission, as annotated in Fig.~\ref{fig:HD163296_v0maps_annotated}. First, there are large positive and negative residuals along the blue-shifted and red-shifted major axes, respectively, labeled ``pressure''. These can be attributed to the sub-Keplerian rotation of the outer disk, driven by the large pressure gradient as the the gas density drops at the edge of the disk \citep{Rosenfeld_ea_2013, Dullemond_ea_2020}. There is also mild contamination from the rear side of the disk, along the top of the plot, spanning between 230\degr{} and 310\degr{}.

We also see three large perturbations, marked by the arc-shaped regions in Fig.~\ref{fig:HD163296_v0maps_annotated} centered at $(r,\, \phi) = (2\farcs85,\, 305\degr)$, $(1\farcs65,\, 315\degr)$ and $(0\farcs9,\, 220\degr)$, labeled A, B and C, respectively. Feature A is the manifestation of the ``kink'' KPS in channel maps reported by \citet{Pinte_ea_2018b}, which the authors attribute to a $\approx 2~M_{\rm Jup}$ planet. The inner most perturbation, C, is coincident with the ``kink'' at 86~au reported by \citet{Pinte_ea_2020}, while the extended arm closely traces the azimuthal asymmetry detected in the mm continuum \citep{Isella_ea_2018} in azimuthal extent, but appears to be radially offset to larger orbital radii by ${\sim} 0\farcs1$. This is discussed more in Section~\ref{sec:discussion:HD163296}. The middle perturbation, B, has no previous detection, but lies at a radius that is broadly consistent with the 145~au ($1\farcs45$) gap in the mm continuum \citep{Isella_ea_2016, Isella_ea_2018}. Both these features appear to have a large azimuthal extent, traced by the dashed lines. In particular, this demonstrates that the previously detected ``kink'' KPS feature is far larger and more complex than previously thought from previous observations, the cause of which is discussed more in Section~\ref{sec:discussion:HD163296}.

In addition to these larger deviations, we find evidence of three arc structures: two well defined arcs in the inner disk at 115 and 181~au ($1\farcs15$ and $1\farcs8$), marked with solid lines labeled D and E, respectively, and one large spiral along the south west edge of the disk at 414~au ($4\farcs1$), shown by the solid line with a faded end, labeled F. As this latter arc blends into the perturbation from the pressure gradient, we are unable to constrain its azimuthal extent.

The $^{13}$CO $v_0$ map, residual from a Keplerian rotation pattern and deprojected residual map are shown in Figs.~\ref{fig:HD163296_v0maps}g, h and i, respectively. Large systematic residuals are seen along the north-west of the disk, similar in morphology to those seen in $^{12}$CO emission, however with much larger velocities. The morphology of these residuals are similar to that proposed by \citet{Yen_Gu_2020} for a misspecified emission surface. As with the $^{12}$CO emission, this arises because the emission starts to become thin in the outer edge of the disk and thus traces both the front and rear side of the disk. The average height traced by the emission is then underestimated by the emission surface, giving rise to these residuals. The pressure gradient in the outer disk also introduces small perturbations along the major axis of the disk, as for $^{12}$CO.

Beyond these features, there is one notable residual: a localized perturbation at a radial offset of $\approx 157$~au ($1\farcs55$), coincident with the large velocity residual annotated as B in Fig.~\ref{fig:HD163296_v0maps_annotated}. Tracing the feature in two isotopologues points towards a real velocity deviation rather than noise. The coincidence of this feature with the gap in the dust continuum, D145, is suggestive of a planetary origin. However, the similar azimuthal angle to the previously discussed planet at $\approx 260$~au, raises the question of whether these features are related to the same origin, a scenario discussed more in Section~\ref{sec:discussion:HD163296}.

Maps showing $v_0$, the residuals from Keplerian rotation and spatially deprojected residuals from Keplerian rotation for C$^{18}$O emission are plotted in Figs.~\ref{fig:HD163296_v0maps}j, k and l, respectively. The residual maps are dominated by large systematic residuals, similar in morphology to those seen in the northern half of the disk traced by $^{12}$CO and $^{13}$CO emission, but also reflected in the southern half of the disk. \citet{Yen_Gu_2020} argue that residuals with this morphology indicate an incorrect inclination. However, for the C$^{18}$O emission this is rather due to emission arising from very close to the midplane such that there is little difference in the brightness of emission arising from the underside of the molecular layer (i.e., from the back side of the disk), or the top side of the molecular layer (i.e., from the front side of the disk). For highly abundant species, such as $^{12}$CO, the difference in gas temperature at the top of the molecular layer (set by the dissociation of CO by UV radiation; $T \gtrsim 50$~K) and the bottom of the molecular layer (set by the freeze-out of CO; $T \sim 21$~K) is sufficiently large that separating emission from these two regions is simple. For less abundant species, such as C$^{18}$O, the top of the molecular layer is much deeper in the disk and thus spans a much narrower temperature range. The result being, disentangling between near-side and far-side emission is much tougher, leading to the line emission tracing an average height much closer to the midplane. This is very similar to the optically thin emission argument made for $^{12}$CO and $^{13}$CO emission above, however we stress that while the outcome is the same in terms of residuals, the reason for the bias in traced height is different.  

\subsection{MWC~480}
\label{sec:rotation_maps:MWC480}

The $v_0$ maps, residuals from Keplerian rotation and spatially deprojected residual maps for CO isotopologue emission from the disk of MWC~480 is shown in Fig.~\ref{fig:MWC480_v0maps}. In all panels, the disk rotation is anti-clockwise. In this source, all velocity deviations are smaller in magnitude to those observed in HD~163296, both in terms of absolute value and relative to the local Keplerian rotation. As with HD~163296, continuum emission from the disk around MWC~480 has been detected out to $\approx 1\farcs7$ meaning that the gas disk, as traced by $^{12}$CO, is up to three times larger in radius. Note that the mm continuum edge reported in MAPS~III \citep{Law_ea_2020_rad}, provides a more conservative estimate of the outer disk as the region which encompasses 90\% of the continuum flux.

\begin{figure}
    \centering
    \includegraphics[width=\columnwidth]{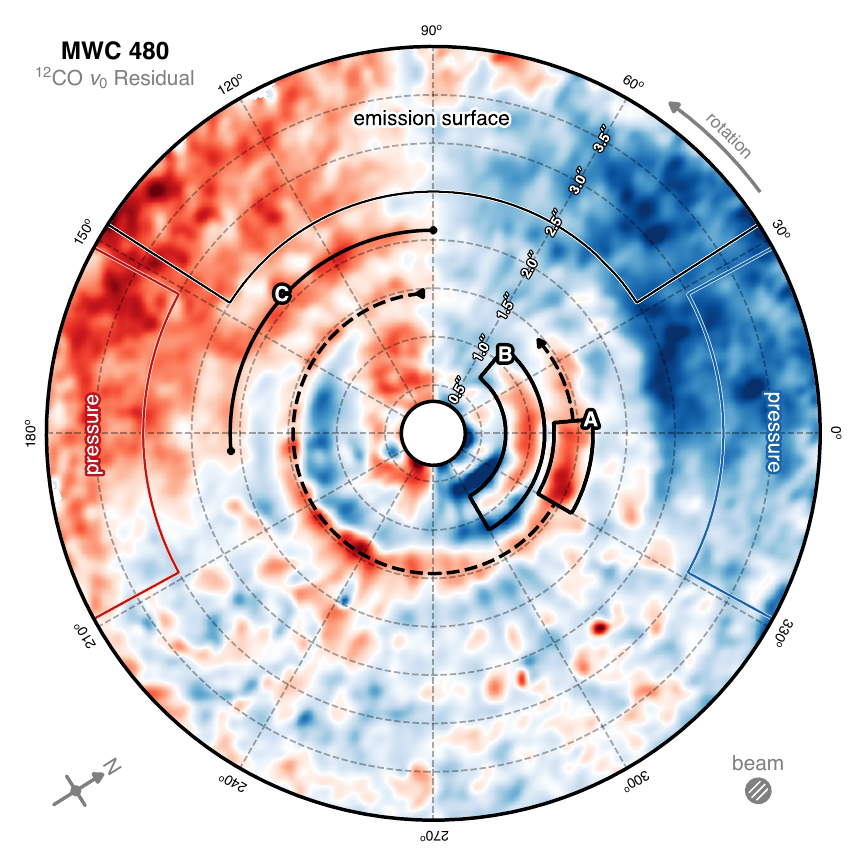}
    \caption{Annotated version of Fig.~\ref{fig:MWC480_v0maps}c, showing the features observed in the residual rotation map from $^{12}$CO in the disk of MWC~480. The disk rotates in an anti-clockwise fashion. Residuals due to the pressure gradient slowing the rotation of the disk are observed along the major axis, marked as ``pressure'', with similar systematic features likely due to a misspecified emission surface marked as ``emission surface''. The most prominent is an arc extending almost the full azimuth of the disk at $1\farcs5$ with a peak marked by a solid annulus, labeled A, with an inner arc at $0\farcs95$, labeled B. An additional tentative arc is annotated as arc C at $2\farcs1$.. A representative beam size is shown in the lower right, however due to the elevated emission surface, the true beam size varies across the map.}
    \label{fig:MWC480_v0maps_annotated}
\end{figure}

As MWC~480 was not a target of DSHARP, we only consider the MAPS data. An annotated version of Fig.~\ref{fig:MWC480_v0maps}c is shown in Fig.~\ref{fig:MWC480_v0maps_annotated}. Similar to HD~163296, there are systematic residuals which can be attributed to large negative pressure gradients, however there is no obvious contamination from the backside of the disk which would show up along the bottom of the panel. Instead, a systematic residual with a positive value in the eastern quarter of the disk and a negative value in the norther quarter (i.e., either side of the minor axis in the NE half of the disk in Fig.~\ref{fig:MWC480_v0maps}b, or either side of 90\degr{} in the top of the disk in Figs.~\ref{fig:MWC480_v0maps}c and \ref{fig:MWC480_v0maps_annotated}) is seen. \citet{Yen_Gu_2020} demonstrate that such a morphology can be attributed to a misspecified emission surface. As discussed in \citet[MAPS~IV]{Law_ea_2020_surf}, the $^{12}$CO emission surface in MWC~480 is only well constrained inwards of 2\arcsec{}, so a misspecified surface is the most likely cause for this feature and annotated as such.

Aside from these systematic structures, the kinematic substructure in MWC~480 is dominated by annular structures with the most striking ring of red-shifted emission at $1\farcs5$, labeled as box A, and then two broad arcs slightly inside and outside this feature at 1\arcsec{} and $2\farcs2$, B and C, respectively. There are noisy features observed at the same radius but extending further in azimuth, suggesting that these arcs may extended over a broader azimuth than annotated, however deeper observations would be required to fully constrain their extent. The velocity perturbations appear to be strongest around $\phi \sim 5\degr$, as marked by the two boxed regions in Fig.~\ref{fig:MWC480_v0maps_annotated}. 

The kinematics traced by $^{13}$CO emission are shown in Figs.~\ref{fig:MWC480_v0maps}d, e and f. There is a substantial difference in the magnitude of the perturbations seen in the velocity traced by $^{13}$CO and those traced by $^{12}$CO emission. There appears to be a subtle positive-to-negative residual across the minor axis of the disk, suggesting a slightly sub-Keplerian rotation of the gas (as discussed with regards to the pressure modulation for $^{12}$CO).

Subtle deviations which appear to trace the same regions as those found in the $^{12}$CO emission are detected, most notably, three arc-like structures in the north-west of the disk at the locations of the peak velocity perturbations in $^{12}$CO. The relationship between these structures and those seen in $^{12}$CO are discussed in Section~\ref{sec:discussion:MWC480}.

The bottom row of Fig.~\ref{fig:MWC480_v0maps} shows the C$^{18}$O $v_0$ map, residuals from Keplerian rotation, and a deprojected map of the residuals, respectively. Little, if any, structure is observed in the residuals. A very low-level `three spur' pattern is observed suggesting a misspecified inclination \citep{Yen_Gu_2020}. As discussed above, and in more detail in Appendix~\ref{sec:app:MWC480_inclination}, the inclination of the MWC~480 disk is hard to constrain, and the lack of residuals seen for $^{12}$CO and $^{13}$CO, which predominantly trace the outer regions of the disk than C$^{18}$O, may suggest a radially varying inclination.

\section{Velocity Profiles}
\label{sec:velocity_profiles}

While a $v_0$ map provides an excellent summary of the dominant velocity fields in a disk, the spectrally resolved nature of the observations provides a wealth of additional dynamical information which is lost when collapsing the data cube. In this section we use the full image cube to derive radial profiles for azimuthally averaged rotational, radial and vertical motions.

\begin{figure*}
    \centering
    \includegraphics[width=\textwidth]{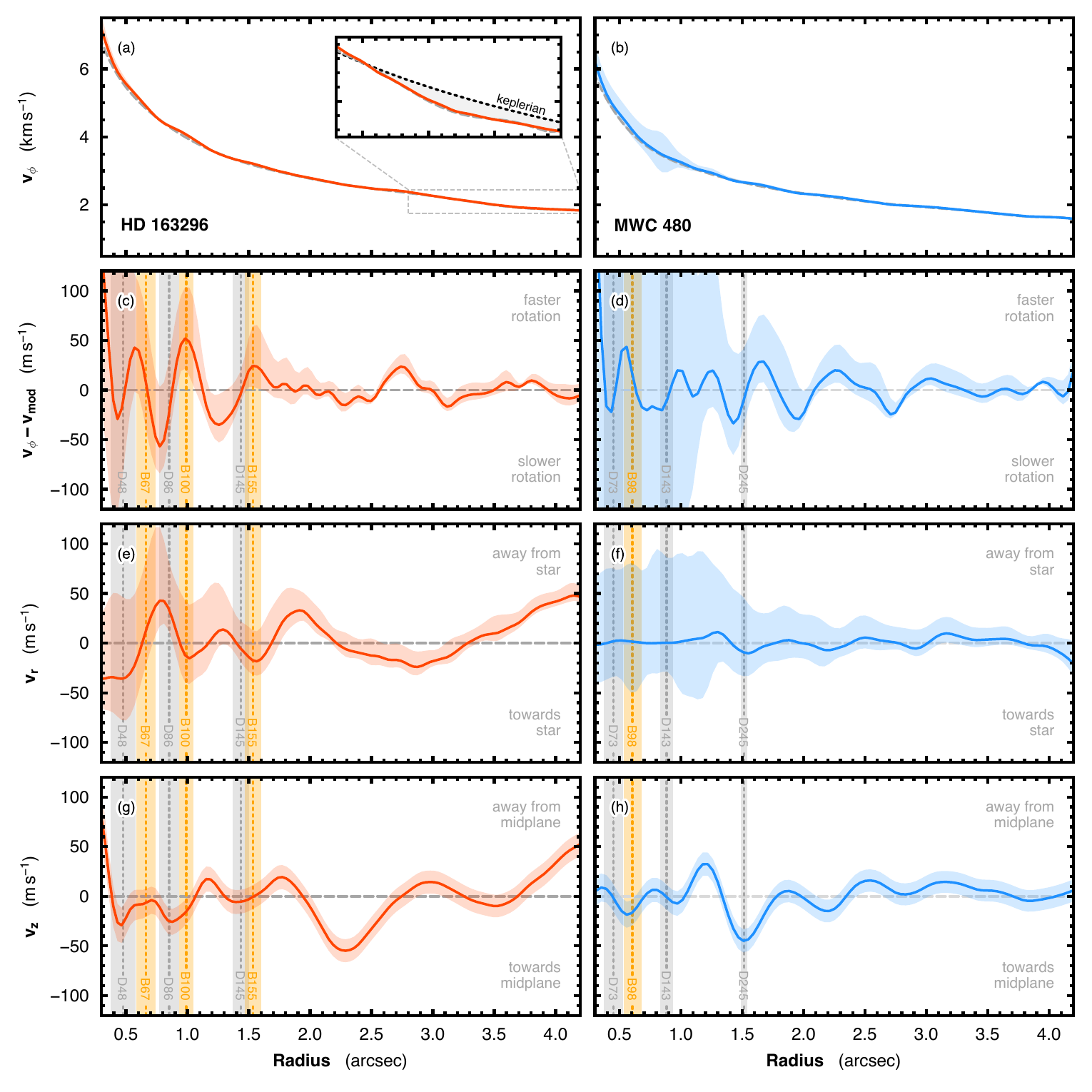}
    \caption{Azimuthally averaged radial profiles of the gas velocities in HD~163296, left, and MWC~480, right, traced by $^{12}$CO. Panels (a) and (b) show $v_{\phi}$, while panels (c) and (d) show the residual of $v_{\rm phi}$ after subtracting a baseline $v_{\rm mod}$. The radial and vertical velocities are shown in panels (e) and (g) and (f) and (h) for HD~163296 and MWC~480, respectively. All velocities have been corrected for the inclination of the disk. Following \citet{Teague_ea_2019b}, the baseline model, $v_{\rm mod}$, used to highlight localized deviation in $v_{\phi}$ is a 10th order polynomial fit to $v_{\phi}$. In (a), an inset highlights the difference between a purely Keplerian rotation profile, black dotted line, and the observed rotation profile and 10th order polynomial fit, gray dashed line, used for panel (c). The shaded regions show the $1\sigma$ uncertainties. Locations of the gaps and rings in the continuum are annotated in the lower three rows as vertical dashed lines with the gray and orange shading showing their width \citep{Huang_ea_2018a}. Note that the sign convention for radial velocity has been inverted relative to \citet{Teague_ea_2019a} such that positive $v_r$ represent gas moving \emph{away} from the star.}
    \label{fig:velocity_profiles}
\end{figure*}

\subsection{Method}
\label{sec:velocity_profiles:method}

To extract the rotational and radial velocities, $v_{\phi}$ and $v_r$, respectively, we use the Python package \texttt{GoFish} \citep{GoFish}. The method is fully described in \citet{Teague_ea_2018c} and extended to three dimensions in \citet{Teague_ea_2019b}. Nonetheless, a brief overview of the method is provided below.

First, the disk is split into concentric annuli with a width of a quarter of the beam major FWHM, taking into account the emission surface for the $^{12}$CO as defined in \citet[MAPS~IV]{Law_ea_2020_surf}. For each annulus, a random sample of spatially independent pixels (i.e., those at least a beam major FWHM apart) are selected, which, under the assumption of an azimuthally symmetric disk, should all be identical in shape, and only differ by their line center which varies due to the projected velocity component of the disk. The velocity axis for each spectrum is then corrected for the disk velocity structure using Eqn.~\ref{eq:v_disk}, assuming $v_z = 0$. The reason for not including a $v_z$ component in this correction is that the projection of this term has no azimuthal dependence and thus only results in an overall shift of the spectral axis making it degenerate with $v_{\rm LSR}$. The correct $v_{\phi}$ and $v_r$ values are defined as those that best align the individual spectra, as quantified by the fit of Gaussian processes mean-model to the averaged spectra. In essence, this minimizes the standard deviation between the sample of shifted spectra without having to make assumptions about the underlying line profile, i.e., assuming an analytical profile. There will be some intrinsic variation in the line profile as a function of azimuth owing to projection effects, however such variations are expected to be negligible compared to thermal noise of the data \citep[see the discussion in][for example]{Teague_Loomis_2020}. 

This process is repeated 20 times for each annulus, each time taking a new, random selection of pixels in order to minimize features in the radial profiles which can be attributed to noisy pixels \citep[e.g.,][]{Keppler_ea_2019}. The posterior distributions from each sample are combined and the results shown in the top three rows of Fig.~\ref{fig:velocity_profiles}. As $v_{\phi}$ is dominated by the Keplerian rotation, a background model is subtracted to highlight the small-scale structure, as in the second row of Fig.~\ref{fig:velocity_profiles}. As discussed in \citet{Teague_ea_2019b}, while the choice of background model does not strongly affect these residuals, different background models will lead to small changes in the overall gradient, and thus the measured `peak-to-trough' values \citep[to measure spatial properties of pressure perturbations, for example;][]{Yun_ea_2019}. Such an approach will only highlight the high spatial frequency (radially localized) perturbations, rather than larger scale perturbations, such as the sub-Keplerian rotation identified in the outer disk of HD~163296. This is demonstrated by the inset axes in Fig.~\ref{fig:velocity_profiles}a which compares a purely Keplerian rotation with a black dotted line to the observed $v_{\phi}$ profile in red and the 10th-order polynomial fit used for Fig.~\ref{fig:velocity_profiles}c, shown by a gray dashed line. As such, these panels should only be used for a qualitative analysis, and any quantitative analysis must be done with the $v_{\phi}$ profiles directly.

To measure the vertical velocity component, an analytical Gaussian fit to the aligned and averaged spectra must be performed in order to recover the line center. The difference between the measured line center and the systemic velocity is subsequently attributed to any vertical velocity components. When aligning and averaging the spectra for each annulus, emission from the back side of the disk was found to contaminate the line wings, resulting in a broad, pedestal-like component. To remove this from the fitting, only velocities within $500~{\rm m\,s^{-1}}$ of the $v_{\rm LSR}$ were considered in the fit which was verified to not include any wing emission. To improve the precision of the fit, the averaged spectrum is binned onto a velocity axis with a channel spacing of $50~{\rm m\,s^{-1}}$, a factor of 4 finer than the native data. This is possible as the velocity shifts due to the the rotation of the disk result in sub-channel shifts, meaning that the average spectrum is sampled at a much higher rate than the channel spacing. As discussed in \citet{Teague_Loomis_2020}, this approach does not remove any systematic broadening effects, for example due to the spectral response of the correlator, but does allow for the overall structure of the line to be more precisely characterized.

It is important to note that while both $v_{\phi}$ and $v_r$ are measured absolutely, $v_z$ is only a relative measurement and depends on the accuracy of the $v_{\rm LSR}$ measurement. For both sources, we estimate the uncertainty on $v_{\rm LSR}$ by fitting $v_0$ maps made from $^{13}$CO and C$^{18}$O emission described in Appendix~\ref{sec:app:rotation_map_fitting}. In both cases, the best-fit $v_{\rm LSR}$ values vary by $\sim\!\!10~{\rm m\,s}^{-1}$, and this scatter is used as the uncertainty on $v_{\rm LSR}$. This uncertainty dominates the uncertainty plotted in panels g and h of Fig.~\ref{fig:velocity_profiles}.

\begin{figure*}
    \centering
    \includegraphics[width=\textwidth]{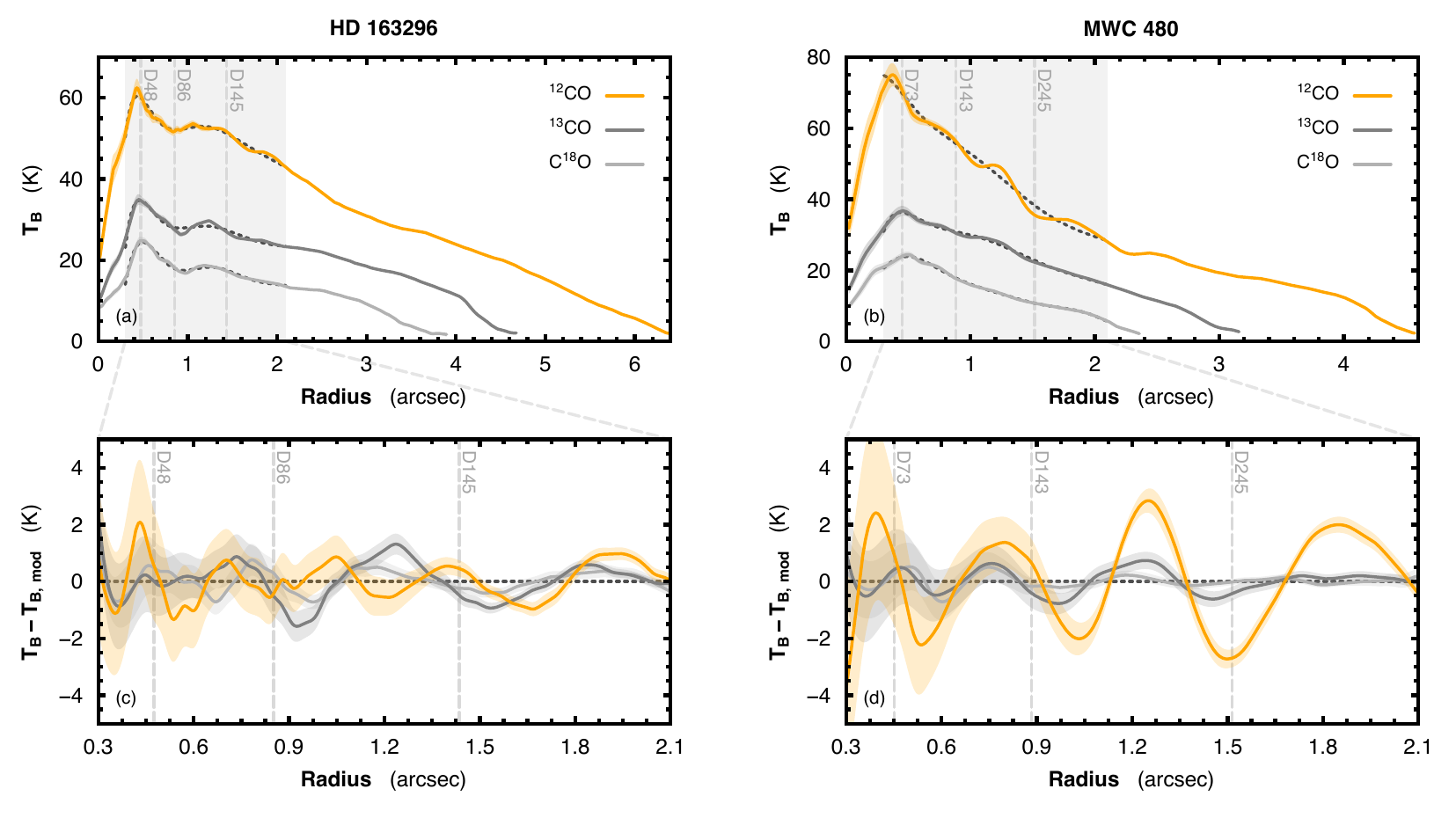}
    \caption{Azimuthally averaged $T_B$ profiles for CO $J = 2-1$ isotopologue emission in HD~163296, panels (a) and (c), and MWC~480, (b) and (d). In all panels, the shaded regions show the $1\sigma$ uncertainty while the vertical dashed lines show the locations of gaps in the mm~continuum. Panels (c) and (d) show a zoom-in of panels (a) and (b), respectively, after subtracting a smooth background model, $T_{B,{\rm mod}}$, shown as a dotted black line in panels (a) and (b). The background model is a 10th-order polynomial fit to the profile between $0\farcs3$ and 4\arcsec{} for $^{12}$CO and $^{13}$CO, and between $0\farcs3$ and 2\arcsec{} for C$^{18}$O. The shaded regions in panels (a) and (b) show the extent of panels (c) and (d), respectively.}
    \label{fig:Tb_profiles}
\end{figure*}

\subsection{HD~163296}
\label{sec:velocity_profiles:HD163296}

The azimuthally averaged velocity profiles traced by $^{12}$CO emission are shown in the left panel of Fig.~\ref{fig:velocity_profiles}. We find that the HD~163296 velocity profiles are consistent with those previously published using the DSHARP data when adopting the same emission surfaces for $^{12}$CO \citep[but noting the swap in sign of $v_r$ to be consistent with the cylindrical coordinate definition;][]{Teague_ea_2018a, Teague_ea_2019a}.

There is a strong correspondence between variations in the gas velocity structure and the continuum gaps and rings. In particular, $v_{\phi}$ will be modulated by changes in the gas pressure gradient due to perturbations in the disk physical structure. Such velocity deviations have been previously used to measure the depth of the gaps \citep[and hence the mass of planets required to open these gaps;][]{Teague_ea_2018a}, and the efficiency of grain trapping in the continuum rings \citep{Rosotti_ea_2020}. Beyond the edge of the mm continuum there are far fewer variations in $v_{\phi}$.

Both radial and vertical components of the velocity, $v_r$ and $v_z$, respectively, appear to have more structure in the outer disk. A radial outflow, previously reported by \citet{Teague_ea_2019a} is recovered, and suggest that $^{12}$CO may be tracing the base of a wind in the outer disk. Intriguingly, HD~163296 hosts jets and a large scale disk wind \citep{Klaassen_ea_2013} which may be associated with these smaller scale dynamical features. \citet[MAPS~XVI]{Booth_ea_2020} present a detailed analysis of the large scale disk wind as seen with MAPS data. In that paper, Booth et al. suggest a wind launch point at $\sim$4~au; in this case it is not clear as to the relationship between the large scale wind and flow arising from the outer disk. The most striking kinematic feature in the outer disk is the large negative $v_z$ component at $r \approx 2\farcs3$ (232~au) associated with meridional flows around an embedded planet \citep{Pinte_ea_2018b, Teague_ea_2019a}, believed to be driving the ``kink'' KPS discussed in Section~\ref{sec:rotation_maps:HD163296}.

\subsection{MWC~480}
\label{sec:velocity_profiles:MWC480}

Velocity profiles for the MWC~480 disk are shown in the right hand column of Fig.~\ref{fig:velocity_profiles} and show a different behavior to those from the HD~163296. Despite the rich substructures observed in the dust continuum for this source \citep[MAPS~XIV]{Long_ea_2018, Liu_ea_2019, Sierra_ea_2020}, there appears to be a much weaker relation between where velocity perturbations are found in the disk and the continuum substructure, however the large uncertainties preclude any definitive statement. In particular, the inner $1\farcs2$ (194~au) appears to dominated by azimuthally asymmetric features, very clearly seen in Fig.~\ref{fig:MWC480_v0maps_annotated}, which result in a poor constraint of $v_{\phi}$ and the resulting large uncertainties on both $v_{\phi}$ and $v_r$. In the outer disk, however, there are significant deviations around $\sim\!1\farcs5$, the same radius as the large arc shown in Fig.~\ref{fig:MWC480_v0maps_annotated}, in addition to potentially related structure extending out to the edge of the disk, with the radial width of the features increasing with radius.

The disk of MWC~480 appears to have negligible $v_r$ components which are azimuthally symmetric, suggesting the absence of a disk wind like HD~163296. However, this does not rule out more localized radial flows, particularly in the inner disk where the uncertainties are large. The largest velocity deviation is seen in the $v_z$ component, tracing a large vertical flow at $r \approx 1.5\arcsec$, seen clearly as the ring in Fig.~\ref{fig:MWC480_v0maps_annotated}. Smaller vertical flows are also seen and correspond to the weak arcs observed in the $v_0$ map residuals.

\section{Gas Temperature}
\label{sec:gas_temperature}

\begin{figure*}
    \centering
    \includegraphics{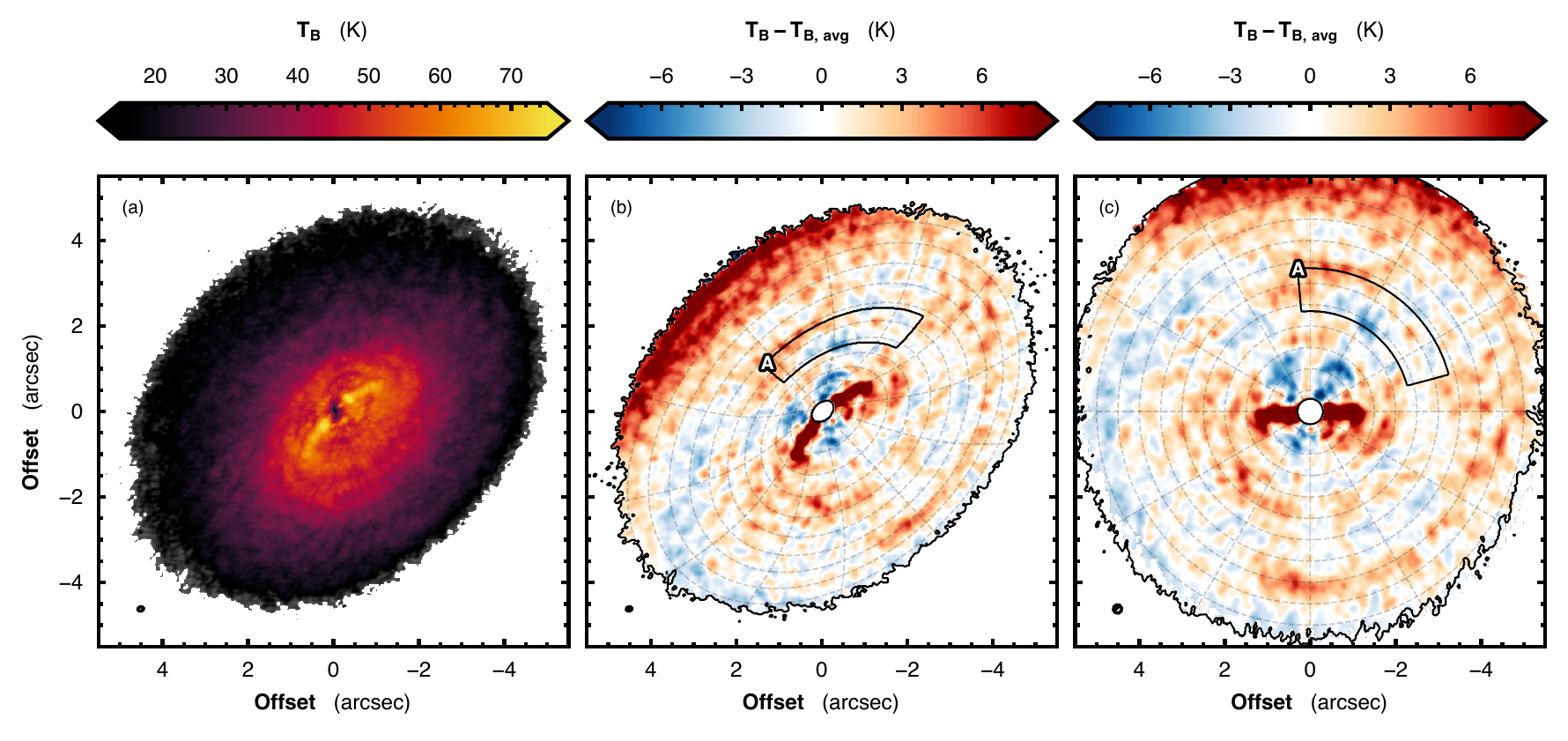}
    \caption{Gas temperature perturbations probed by $^{12}$CO in HD~163296. The panel (a) shows the $T_B$ map, tracing the gas temperature of the top surface. Panel (b) shows the residual after subtracting the azimuthally averaged $T_B$ profile. Panel (c) shows the same residuals, but deprojected into a face-on orientation. A representative beam size is shown in the lower right, however due to the elevated emission surface, the true beam size varies across the map. In panels (b) and (c), dashed contours show lines of constant radius and azimuth in steps of $0\farcs5$ and 30\degr{}. In all frames, the disk rotates in clockwise direction. The region where \citet{Pinte_ea_2018a} detected a KPS, label box A in Fig.~\ref{fig:HD163296_v0maps_annotated}, is shown in both (b) and (c).} 
    \label{fig:HD163296_12CO_Tbresiduals}
\end{figure*}

For optically thick emission, such as $^{12}$CO, we can use the brightness temperature, $T_B$, of the emission as a proxy for the gas temperature. This has been previously suggested as a method to detect embedded planets via their heating of their immediate surroundings \citep{Cleeves_ea_2015}, with recent studies finding spiral structure likely associated with possible planets \citep[e.g.,][]{Teague_ea_2019b, Wolfer_ea_2020}. In this section, we look for variations in the temperature structures probed by CO isotopologue emission which correlate with the velocity structures we have detected in the disks of HD~163296 and MWC~480.

\subsection{Method}
\label{sec:gas_temperature:method}

To measure an azimuthally averaged $T_B$ profile for each disk, we use the $T_B$ maps produced by \texttt{bettermoments} when calculating the $v_0$ maps described in Section~\ref{sec:rotation_maps}. Using \texttt{GoFish}, these maps were binned into annuli with a width of 1/4 of the beam major axis, taking into account the elevated emission surface of $^{12}$CO by using the analytical fits described in \citet[MAPS~IV]{Law_ea_2020_surf}, and then averaged. To estimate the uncertainty, the standard deviation of each annular bin was divided by the square root of the number of independent beams which fit in that bin. These were converted to units of Kelvin using the full Planck law. The resulting profiles are shown in Fig.~\ref{fig:Tb_profiles}, with the gaps in the continuum emission included. The inner $0.4\arcsec$ of all profiles are affected by beam dilution and continuum subtraction, resulting in a drop in $T_B$. Beam dilution will occur at radial separations larger than the beam as the projected Keplerian rotation leads to a narrow azimuthal extent of the emission which remains unresolved \citep{Horne_Marsh_1986}.

The radial profiles show a level of substructure that is not as clearly seen in the radial profiles of the integrated intensities for strong lines \citep[aside from HCO$^+$, see][MAPS~III]{Law_ea_2020_rad}. This is because while integrated intensity and peak brightness are related for a single emission source, the back side of the disk will contribute to the integrated intensity in a spatially dependent manner, while the peak brightness temperature should only be affected by the front side emission (at least for optically thick lines). 

\begin{figure*}
    \centering
    \includegraphics{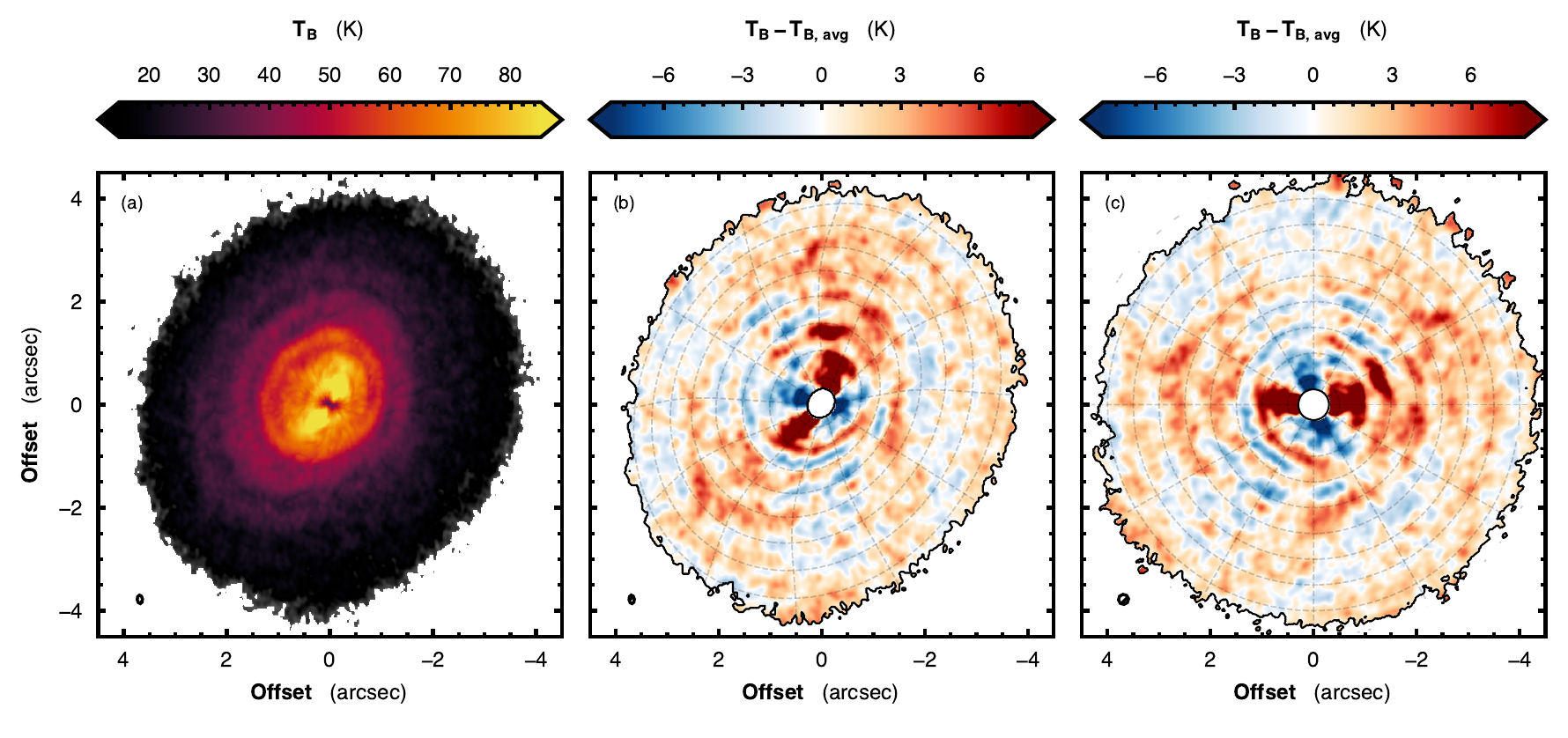}
    \caption{As Fig.~\ref{fig:HD163296_12CO_Tbresiduals} but for $^{12}$CO in the disk of MWC~480. In all frames, the disk rotation is in an anti-clockwise direction.}
    \label{fig:MWC480_12CO_Tbresiduals}
\end{figure*}

These azimuthally averaged profiles are used to create a 2D background model, which can be subtracted from the $T_B$ maps to reveal subtle perturbations in the background temperature structure \citep[e.g.,][]{Teague_ea_2019a}. Figures~\ref{fig:HD163296_12CO_Tbresiduals} and \ref{fig:MWC480_12CO_Tbresiduals} show, for HD~163296 and MWC~480, respectively, the $T_B$ map in panel (a), the residual after subtracting the azimuthally averaged model, $T_{B,\,{\rm avg}}$ in panel (b), and the residuals deprojected to account for the inclination and projection of the emission surface in panel (c). For both sources, an enhanced $T_B$ is found along the major axis of the disk extending $\sim\! 1\arcsec$ from the disk center. This arises due to the Keplerian rotation pattern. Gas tracing smaller radii, which is therefore hotter and brighter, is detected primarily along the major axis at small radial offsets due to the projection effects from the inclination of the disk \citep[see Fig.~1 in][for example]{Horne_Marsh_1986}.

\subsection{HD~163296}
\label{sec:gas_temperature:HD163296}

The radial $T_B$ profiles for HD~163296 are shown in the left column of Fig.~\ref{fig:Tb_profiles}. Small deviations relative to the mean background are observed in all three isotopologues, likely associated with the continuum substructure in the inner $1\farcs5$. However, the outer disk appears fairly featureless. The residuals of the $T_B$ profiles for the three CO isotopologues after subtracting a background model (a 10th order polynomial fit, as done in Section~\ref{sec:rotation_maps:method}), are shown in Fig.~\ref{fig:Tb_profiles}c. However, the radial variations appear to be associated with the ring and gap structure, and point towards variations in the thermal structure of the disk driven by changes in the dust opacity in gapped regions \citep[e.g.,][]{Teague_ea_2017, Facchini_ea_2018, vanderMarel_ea_2018, Alarcon_ea_2020, Rab_ea_2020}. It is important to note that optically thick continuum emission has been shown to impact the emergent radiation, either through the absorption of emission arising from behind it, or through continuum subtraction \citep[e.g.,][]{Boehler_ea_2018, Weaver_ea_2018}, potentially introducing spurious features in molecular line emission. This is unlikely the case here as the same radial structure is seen when using data with and without continuum subtraction. Furthermore, with the molecular emission arising from an elevated surface, the continuum emission does not trace the same annuli when projected on the sky, limiting the impact of the absorption.

There are subtle substructures seen in the $T_B$ residual map, shown in Fig.~\ref{fig:HD163296_12CO_Tbresiduals}b. A tentative arc is seen along the south-western quadrant at a radius of $\sim\!3\farcs5$. With the current sensitivity it is hard to constrain its azimuthal extent, however it appears to have an annular rather than a spiral morphology. There is only tentative structure observed around the location of the planet, at ($0\farcs0,\, 1\farcs5$) in central panel and (-$1\farcs5,\,2\farcs5$) in the projected panel. The ``kink'' KPS feature seen in Fig.~\ref{fig:HD163296_v0maps}b is bounded to the east (thus trailing the hypothesized planet) by slightly hotter gas, while at the location of the planet we find a slightly lower gas temperature. Deeper observations, facilitating a comparable mapping of the $^{13}$CO and C$^{18}$O emission, are required to fully map the temperature structure around the planet and thus quantify the significance of any localized heating from accretion onto the planet.

\subsection{MWC~480}
\label{sec:gas_temperature:MWC480}

In contrast, MWC~480 displays a more complex range of temperature profiles with ringed features seen across the inner 3\arcsec{} of the disk, as shown in Fig.~\ref{fig:Tb_profiles}b. The substructures seen in the $T_B$ profiles, Fig.~\ref{fig:Tb_profiles}d, appear not to strongly correlate with the location of the continuum substructures, with the exception of the large dip in $^{12}$CO seen at 245~au. Of particular note is that the features traced by different isotopologues are not merely offset in radius, which could be explained by an incorrect emission surface used in the projection, but they appear to have different radial scale wavelengths such that their peak-to-peak distances increases with height traced in the disk. This phenomenon is discussed more in Section~\ref{sec:discussion:MWC480}.

\begin{figure}[t]
    \centering
    \includegraphics[width=\columnwidth]{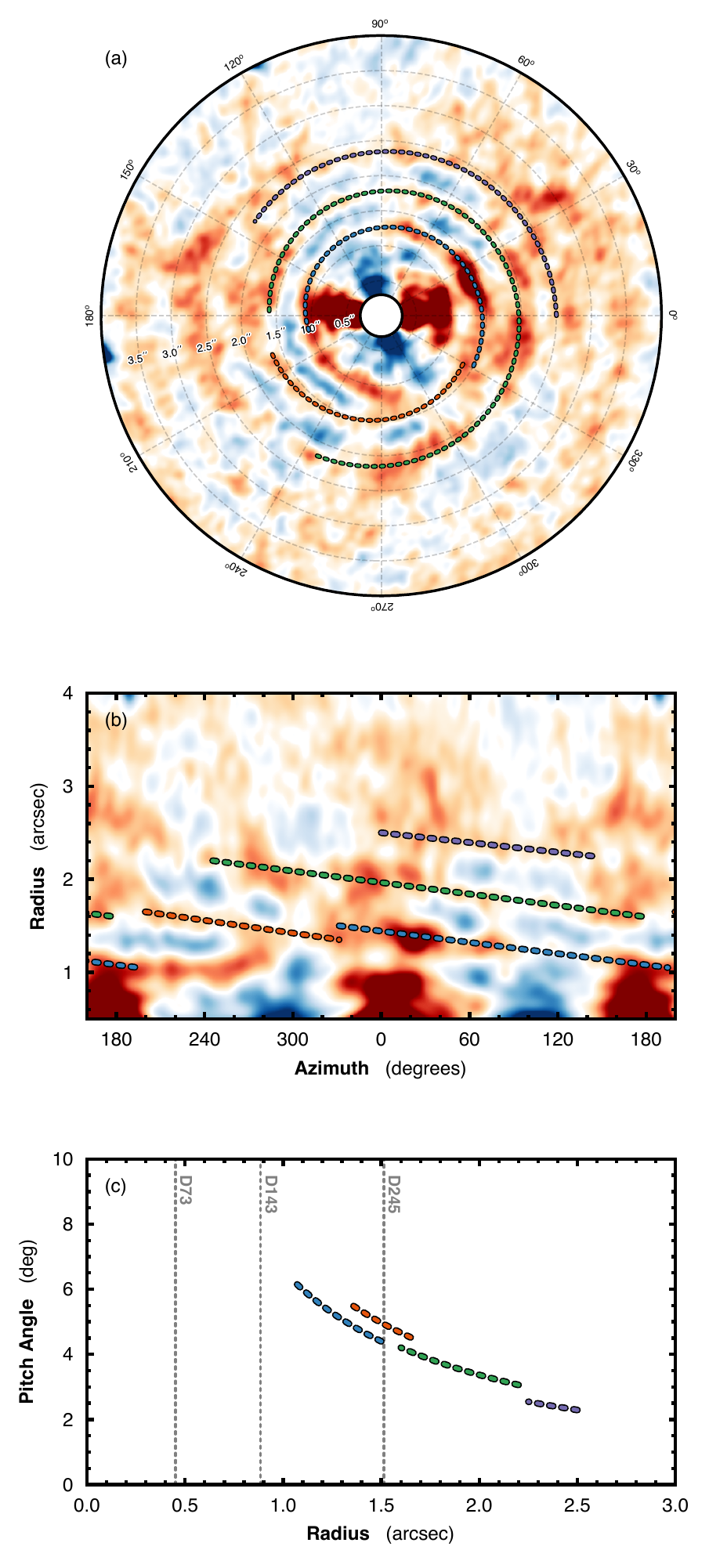}
    \caption{Panel (a) shows an annotated version of Fig.~\ref{fig:MWC480_12CO_Tbresiduals}c, highlighting the spirals, show with colored dotted lines. Panel (b) is a polar deprojection of panel (a) to better demonstrate the spiral structure. The pitch angles for the spirals are shown in panel (c), with the locations of the continuum gaps annotated.}
    \label{fig:MWC480_12CO_spirals}
\end{figure}

MWC~480 displays tentative substructure when looking at the full 2D map, characterized by multiple, tightly wound spirals. The top panel of Figure~\ref{fig:MWC480_12CO_spirals} shows an annotated version of Fig.~\ref{fig:MWC480_12CO_Tbresiduals}c, with the spiral arms shown by colored dotted lines. The spirals are only identified by-eye, with the quality of the data precluding a more quantitative fit. Deeper observations are required to more accurately map these subtle $T_B$ variations. Figure~\ref{fig:MWC480_12CO_Tbresiduals}b shows a polar deprojection of this panel. These spirals features align with the two spirals observed in the gas velocity (see Fig.~\ref{fig:MWC480_v0maps_annotated}), where the temperature perturbations are found to be both leading and trailing the velocity perturbations. We find no analogous perturbation in the velocity structure to the outer spirals. The pitch angle of the spirals, $\beta$, are defined as $\tan(\beta) = -{\rm d}r / r {\rm d}\phi$ and plotted in the bottom panel of  Fig.~\ref{fig:MWC480_12CO_spirals}, with the gaps in the mm continuum shown for reference. The significance of the radial dependence on $\beta$ is discussed in Section~\ref{sec:discussion:MWC480:planet}.

\section{Discussion}
\label{sec:discussion}

We have shown that the disks around both HD~163926 and MWC~480 exhibit a substantial amount of substructure in the gas traced by $^{12}$CO emission, both in the gas velocity structure and temperature structure. These structures are predominantly at radii beyond the continuum edge of the disks, demonstrating that either the processes that are sculpting the large grains traced by \mbox{(sub-)mm} continuum emission also influence the outer disk, or that there are additional processes present in the outer disk. In this section, we interpret the observed substructure as the signatures of embedded planets.

\subsection{HD~163296}
\label{sec:discussion:HD163296}

With these data, we are able to confirm many of the previously detected kinematic signatures of embedded planets in HD~163296. In particular, we recover the `kink' reported by \citet{Pinte_ea_2018b} at the outer edge of the disk, finding that it is much more azimuthally extended than previously inferred. In addition, we find azimuthally averaged velocity profiles which agree with those previously reported \citep{Teague_ea_2018a, Teague_ea_2019b}, which showed pressure-modulated rotational velocities due to perturbations in the gas surface density \citep{Rosotti_ea_2020}, and localized radially converging flows, interpreted as the tops of meridional flows \citep{Szulagyi_ea_2014, Morbidelli_ea_2014}.

\subsubsection{A Planet at 260~au}

With observations at higher spectral resolution and achieving a higher sensitivity, the KPS previously detected (`kink' in the emission morphology) by \citet{Pinte_ea_2018b} is clearly seen in the rotation map residuals (Fig.~\ref{fig:HD163296_v0maps_annotated}, feature A), and associated with a much more azimuthally extended structure. This is also reflected in the range of velocities (i.e., number of channels) the feature appears in channel maps, shown in Appendix~\ref{sec:app:channel_maps}, where the perturbation is seen to change morphology as a function of velocity.

As these residuals extended over large azimuthal ranges, it is likely that at large azimuthal extends from the planet there is little azimuthal dependence on their projection and thus that the velocities in these regions are dominated by vertical flows \citep[Eqn.~\ref{eq:v_disk}, and see the discussion in][]{Teague_ea_2019a}. Although \citet{Pinte_ea_2018b} only originally considered perturbations in $v_{\phi}$ when modeling HD~163296, \citet{Pinte_ea_2019} reported more numerical simulations that showed that embedded planets drive perturbations in all three directions, with vertical motions starting to become present at higher altitudes, while radial and rotational perturbations are stronger in midplane regions. \citet{Bae_ea_2021} also showed that when using more realistic thermal structures for the background disk, planets will excite buoyancy resonances, driving tightly wound spirals with large vertical motions dominating the dynamics. A potential scenario is that perturbations traced by $^{12}$CO should be dominated by in-plane motions, $v_{\phi}$ and $v_r$, close to the planet, and predominantly in the vertical direction at larger azimuthal offsets from the planet.

However, \citet{Bae_ea_2021} also predict that for buoyancy resonances, the velocity deviations relative to the background Keplerian rotation will be largest at the location of the planet. This is consistent with observed velocity deviations and the predicted location of the planet proposed by \citet{Pinte_ea_2018b}. For HD~163296, unfortunately this location is also where the projected components of $v_{\phi}$ and $v_r$ can add coherently ($\sin(\phi) \approx \cos(\phi)$, so $\phi \approx 45\degr$ or $225\degr$, see Eqn.~\ref{eq:v_disk}), meaning we have the greatest sensitivity to these motions at this particular location. Detecting how the velocity perturbations vary as a function of height and comparing with numerical simulation based predictions may help distinguish between spiral launching scenarios. However, the only true way to distinguish scenarios will require the orbital rotation of the planet such that the projection of the velocity components change significantly. At 230~au the orbital velocity around a $2~M_{\odot}$ star is $\sim\! 5.8~{\rm mas~year^{-1}}$, meaning that with current instrumentation it will be around a decade until a discernible difference can be detected (the limiting factor here being how well a peak or point of reference can be determined in each image).

A planet can readily account for the observed substructure in the disk of HD~163296, without the need for the fine-tuning of any of the disk properties \citep[e.g.,][]{Pinte_ea_2018b, Teague_ea_2018a}. A full comparison with numerical simulations of planet-disk interactions will be the focus of a future article. The inference of an embedded planet associated with meridional flows represents a unique opportunity to probe the delivery of volatile materials to the still-forming atmosphere. Future observations that achieve a much greater sensitivity are required if we are to compare the chemical complexity observed in the inner regions of disk to that in the immediate vicinity of wide separation, embedded planets.

\subsubsection{Additional Localized Velocity Deviations}

\paragraph{A Second Outer Planet?}

In Section~\ref{sec:rotation_maps:HD163296}, we showed the presence of a localized velocity deviation broadly coincident in radius with the D145 gap in continuum emission (feature B in Fig.~\ref{fig:HD163296_v0maps_annotated}). Applying the same method of $v_0$ map making and subtracting a Keplerian velocity model to the $^{13}$CO (2-1) data, we find a similar velocity deviation at the same location, as shown in Fig.~\ref{fig:HD163296_v0maps}. The lower signal-to-noise ratios of the $^{13}$CO images results in a residual map that is far nosier, but the coincidence in location with the $^{12}$CO feature strongly suggests this is a real perturbation, and has a vertical extent which spans between the layers probed by $^{12}$CO and $^{13}$CO emission: $z/r \approx 0.3$ and 0.15, respectively \citep[MAPS~IV]{Law_ea_2020_surf}. It is suspicious, however, that this feature shares the same azimuthal angle as that of the planet at 260~au, and could suggest that a large planet can generate kinematic features across a large radial extent. Simulations of planet-disk interactions employing the most realistic thermal and density structures will be essential in distinguishing between a second planet, or perturbations from a single planet influencing a much more extended radial range than previously thought \citep[e.g.,][]{Dong_ea_2017, Bae_ea_2018a, Bae_ea_2018b}.

\begin{figure}
    \centering
    \includegraphics[width=\columnwidth]{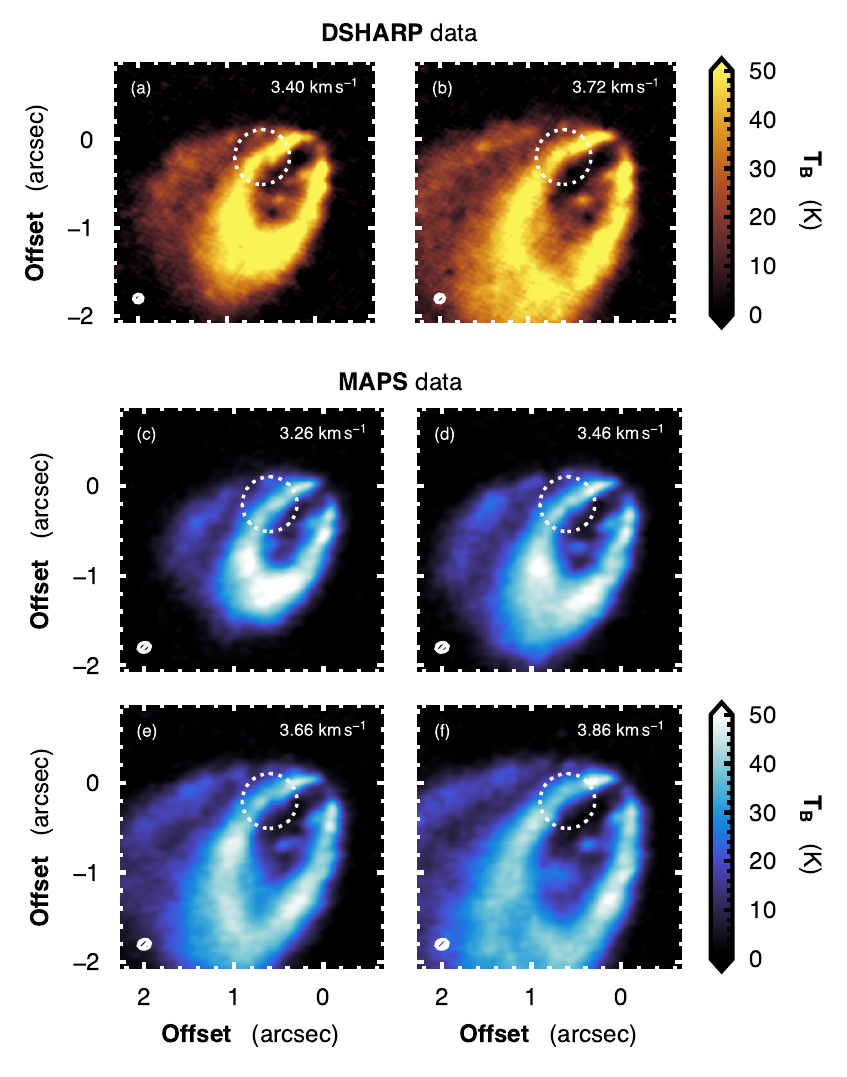}
    \caption{Channel maps of $^{12}$CO $J = 2-1$ emission from the disk around HD~163296, centered on the proposed kinematic planetary signature reported by \citet{Pinte_ea_2020}. The top two panels, (a) and (b) used DSHARP data \citep{Andrews_ea_2018, Isella_ea_2018}, while the bottom four panels, (c) through (f), use MAPS data. No clear ``kink'' is visible at the spatial resolution of the MAPS data. The dotted circle shows the region of interest.}
    \label{fig:HD163296_channelmaps_pinte}
\end{figure}

\paragraph{A Secondary KPS in the Inner Disk}

\begin{figure*}
    \centering
    \includegraphics[width=\textwidth]{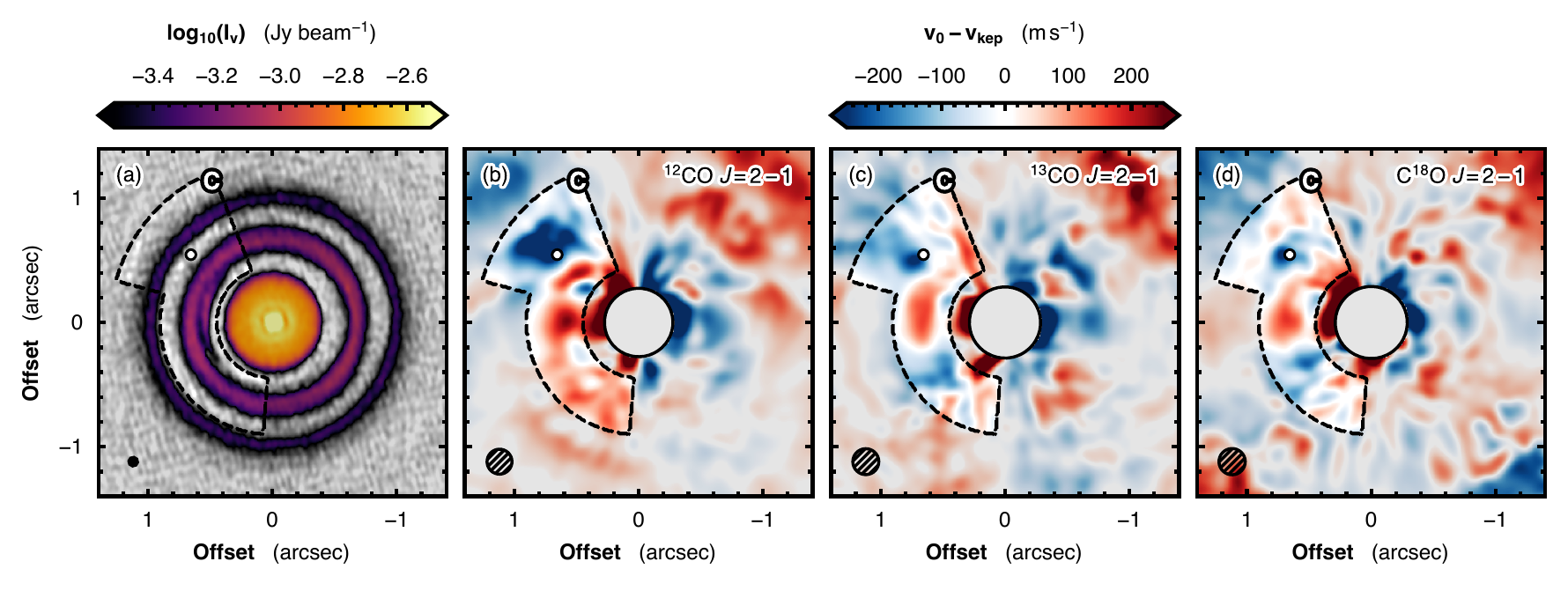}
    \caption{Kinematic structures around the continuum vortex and proposed KPS in the inner disk of HD~163296. Panel (a) shows the spatially deprojected 1.25~mm continuum data from DSHARP \citep{Andrews_ea_2018, Isella_ea_2018} with the vortex and KPS region, box C in Fig.~\ref{fig:HD163296_v0maps_annotated}, outlined. Panels (b), (c) and (d) show similar zoom-ins of the $v_0$ residuals for $^{12}$CO, $^{13}$CO and C$^{18}$O, respectively. In all panels, the proposed planet location from \citet{Pinte_ea_2020} is plotted. All $v_0$ residual maps share the same color scaling. A representative beam size is shown in the bottom left of each panel, taking into account the deprojection.}
    \label{fig:vortex_kinematics}
\end{figure*}

\citet{Pinte_ea_2020} identified a second ``kink'' KPS in the DSHARP set of images. The authors argued that if this feature was driven by an embedded planet, it would require the planet to sit within the D86 gap. Figure~\ref{fig:HD163296_channelmaps_pinte} shows zoomed-in channel maps of the $^{12}$CO emission at the same location as the proposed velocity deviation and in comparison with the DSHARP data \citep{Andrews_ea_2018, Isella_ea_2018}. As the spatial resolution of the MAPS data is slightly coarser than that achieved by the DSHARP observations, we are unable to verify the feature in the channel maps. However, residuals in the $^{12}$CO $v_0$ maps made with both the MAPS and DSHARP data, are found at the proposed location of the planet at $\approx 86$~au, as seen in Fig.~\ref{fig:vortex_kinematics}b, in addition to far more tentative features seen in the $^{13}$CO and C$^{18}$O $v_0$ maps, shown in Fig.~\ref{fig:vortex_kinematics}c,d. A blue / red feature is seen around the proposed planet location, reminiscent of the ``Doppler flip'' found in HD~100546 \citep{Casassus_Perez_2019}, but the orientation of the feature appears inconsistent with the clockwise rotation of the disk.

This ``kink'' KPS also lies very close to the azimuthal asymmetry detected in the continuum emission at $r \approx 0\farcs54$ \citep[55~au, Fig.~\ref{fig:vortex_kinematics}a {and feature C in Fig.~\ref{fig:MWC480_v0maps_annotated}};][]{Isella_ea_2018}. It is unknown what would cause such a feature, however it would require a local pressure maximum to radially and azimuthally capture the large grains. Recent work looking at the kinematics around the vortices, particularly in HD~142527, have suggested that such a pressure maximum has observable effects on the local dynamical structure of the disk \citep{Huang_Isella_ea_2018}. For the dust trap in HD~142527, this manifested as super-Keplerian rotation along the outer edge of the feature \citep[a positive $v_0$ residual;][]{Yen_Gu_2020, Boehler_ea_2021}, consistent with the velocity residuals seen here.

The features in the $v_0$ map residuals support the hypothesis of localized velocity deviations. However, given the spatial resolution, it is hard to distinguish between the source of the velocity deviations and those potentially arising from radiative transfer effects due to the large intensity gradients associated with gaps in the continuum \citep[e.g.,][]{Keppler_ea_2019, Rab_ea_2020, Boehler_ea_2021}. Follow-up observations achieving spatial resolutions of $< 0\farcs1$ are absolutely required if we are to probe the velocity structure of gas on spatial scales equivalent to the substructure observed in the dust continuum.

\subsection{MWC~480}
\label{sec:discussion:MWC480}

MWC~480 exhibits an extensive variety of substructures as traced by the $^{12}$CO emission, both in the velocity as well as the temperature. Within the inner $\sim\!1\farcs2$, there are substantial velocity perturbations with narrow azimuthal extents which preclude a reasonable inference of the $v_{\phi}$ and $v_r$ radial profiles, as evidenced by the large uncertainties for these profiles, shown in Fig.~\ref{fig:velocity_profiles}. Outside this radius, there are significant deviations in $v_{\phi}$ and $v_z$, while perturbations in the radial velocity appear to be minimal. In turn, these velocity perturbations are associated with large-scale variations in the gas temperature, a combination of annular structures which have variations on the order of $\approx 3$~K ($\approx 5\%$ relative to the background), with spiral perturbations detected on top of these, also with variations around $\approx 3$~K. As the disk rotates anti-clockwise on the sky, these spirals are consistent with trailing spirals. 

\subsubsection{Ordered Substructure}
\label{sec:discussion:MWC480:substructure}

What is particularly striking about the features in MWC~480 is how ordered the deviations are: features appear incredibly regular as a function of radius, particularly when compared to the structure observed in the disk of HD~163296 (see Figs.~\ref{fig:velocity_profiles} and \ref{fig:Tb_profiles}). To measure the size of these features, we use a Savitzky-Golay filter (with a window size of $0\farcs45$ and a 3rd-order polynomial fit) to extract the second derivative of the $v_{\phi}$ and $T_B$ profiles. This approach was found to yield a good compromise between smoothing the data to suppress noise and obtaining an accurate measure of radial location of features in the profile. We identify peaks and troughs in the profiles as local minima and maxima in the second-derivative profiles, respectively. The radial width of these features, either the peak-to-peak or trough-to-trough distance, and their radial location, are plotted in Fig.~\ref{fig:MWC480_profile_features}. A linear fit to each of these is shown by the dashed lines, finding a radial depends of ${\sim}\,0.39r$ for the gas velocity and ${\sim}\,0.15r$ for the gas temperature.

It is well known that hydrodynamical processes that create structure in disks will result in features which scale with the local pressure scale height, i.e., increasing with radius. This is routinely seen where the width of gaps and rings formed by a planet with a given mass scales with the local scale height \citep[e.g.,][]{Kanagawa_ea_2015b, Zhang_ea_2018, Yun_ea_2019}, which in turn increases with radius. The strong correlation shown in Fig.~\ref{fig:MWC480_profile_features} between the size of features detected in both the gas temperature and velocity structure and their orbital radius suggest that there appears to be a dominant hydrodynamical process sculpting the structure of the disk.

This is counter to the scenario observed in HD~163296, where the gaseous substructures appear to be somewhat independent of one another. A possible cause could be the difference in the number of planets in each disk. HD~163296 is believed to host (at least) 3 Jupiter-mass planets (inferred through gas and dust surface density depletions, \citealt{Isella_ea_2016}, and through gas dynamics \citealt{Pinte_ea_2018b, Teague_ea_2018a}), such that the influence of each planet will be felt most strongly in radial regions close to the perturber. If MWC~480 were to only host a single planet, a scenario exploring in the next subsection, then the observed deviations would be characterized only by the single perturber.

\subsubsection{A Possible Planet at 245~au}
\label{sec:discussion:MWC480:planet}

\begin{figure}[t]
    \centering
    \includegraphics[width=\columnwidth]{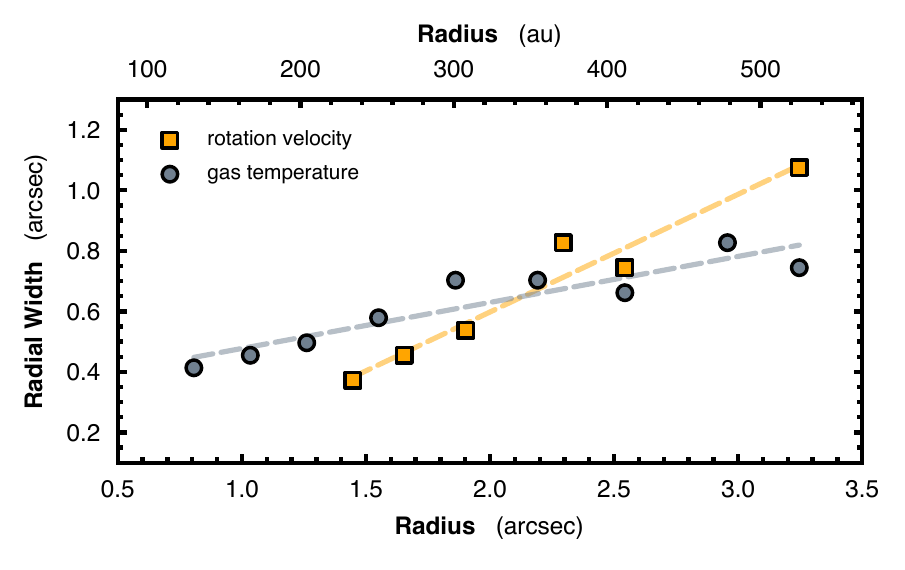}
    \caption{Radial dependence on the size of gaseous substructures in the disk of MWC~480. The figure compares the radial size of the perturbations (maximum-to-maximum or minimum-to-minimum of the second derivative of the radial profile) as a function of radius. The dashed lines show a linear fit to each set of points.}
    \label{fig:MWC480_profile_features}
\end{figure}

An embedded planet would provide an enticing mechanism to explain the observed substructures in the gas and dust. To summarize, the main observed features in the gas are:

\begin{itemize}
    \item Two tightly wound spiral structures seen in the $v_0$ residuals at 162~au ($1\arcsec$) and 245~au ($1\farcs5$). See Fig.~\ref{fig:MWC480_v0maps_annotated}.

    \item Rings of large ($\approx\! 50~{\rm m\,s^{-1}}$) vertical velocities, with the most prominent at 245~au ($1\farcs5$; feature A in Fig.~\ref{fig:MWC480_v0maps_annotated}). See Figs.~\ref{fig:MWC480_v0maps_annotated} and \ref{fig:velocity_profiles}.

    \item Tightly wound trailing spirals in the gas temperatures, with all pitch angles increasing with smaller radii. See Fig.~\ref{fig:MWC480_12CO_spirals}.

    \item A series of temperature minima and maxima extending across the radius of the disk, with the radial width of the features increasing with radius. See Fig.~\ref{fig:MWC480_profile_features}.
\end{itemize}

\begin{figure}[t]
    \centering
    \includegraphics[width=\columnwidth]{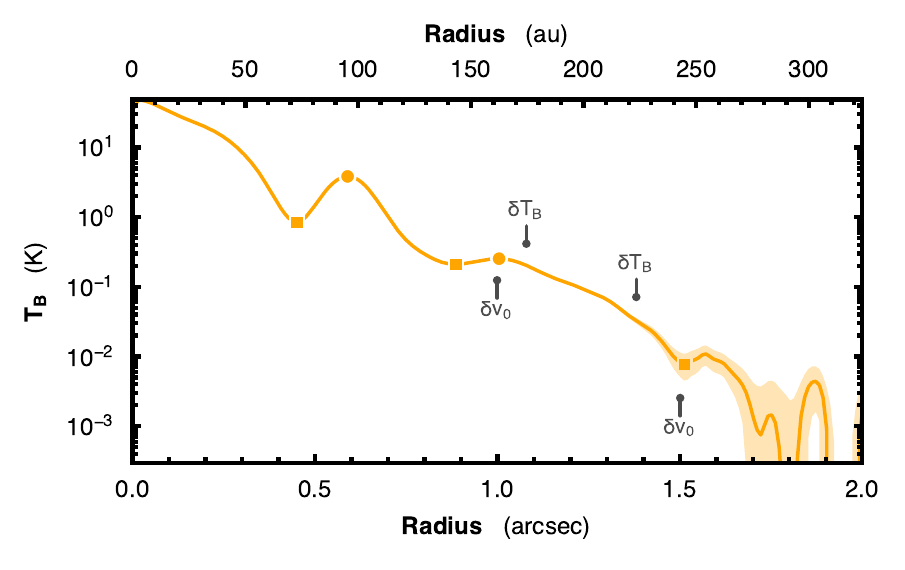}
    \caption{Azimuthally averaged 242~GHz continuum $T_B$ profile for MWC~480 \citep{Sierra_ea_2020}. The shaded region shows the $1\sigma$ uncertainty, estimated by the standard deviation in each radial bin. Three gap centers, D73, D143 and D245 are marked by orange squares, and two rings, B95 and B163, are marked by orange circles. The location of the two annular structures in $T_B$ are marked by $\delta T_B$ (see Fig.~\ref{fig:MWC480_12CO_spirals}), while the two dominant spirals in $v_{\phi}$ (see Fig.~\ref{fig:MWC480_v0maps_annotated}) are marked as $\delta v_{\phi}$.}
    \label{fig:MWC480_continuum}
\end{figure}

\noindent In addition to the complex structures in the gas, there are gaps detected in the mm continuum distribution \citep{Law_ea_2020_rad, Sierra_ea_2020}, as shown in Fig.~\ref{fig:MWC480_continuum}. Extending the list of features, we include:

\begin{itemize}
    \item Two gaps at 73~au ($0\farcs46$) and 141~au ($0\farcs87$), with the latter being a broader structure \citep{Long_ea_2018, Law_ea_2020_rad}. See Fig.~\ref{fig:MWC480_continuum}. 
\end{itemize}

In the following, we identify the most likely location for a planet able to account for these observed features. Without the use of numerical simulations and radiative transfer modeling, we cannot say whether the planet can drive all the observed features, or only a subset of them, however that is the focus of a future paper. The aim of this section is to qualitatively understand which features are consistent with a planet, and where that planet would need to be located.

\paragraph{Gaps in the Continuum}

Any planet of sufficient mass to drive the substructure observed in the gas would also be massive enough to open a gap in the continuum \citep[e.g.,][]{Papaloizou_Lin_1984, Paardekooper_Mellema_2004}. This suggests that the planet could sit within one of the two continuum gaps, 73~au or 143~au, or outside the mm continuum at $\gtrsim 290$~au. There is a tentative dip in the continuum at 243~au ($1\farcs5$), as shown in Fig.~\ref{fig:MWC480_continuum}, however deeper and higher angular resolution observations are needed to verify the presence of this feature. Note that this feature is not reported in MAPS~III \citep{Law_ea_2020_rad} due to the feature lying outside the radius which encloses 90\% of the disk flux which was taken as the disk edge. To distinguish between these potential locations, we must look to the gaseous substructures.


\paragraph{Spirals}

There are two mechanisms that can launch spiral arms from an embedded planet: the traditional Lindblad spirals \citep{Goldreich_Tremaine_1979, Ogilvie_Lubow_2002}, and the more recently discussed spirals arising from buoyancy resonances \citep{Zhu_ea_2012, Bae_ea_2021}. The former drives spirals that have a maximum pitch angle at the location of the planet, that is, at larger radial separations from the planet, the spirals become more tightly wound. In addition, these are usually large pitch angles, reaching $\gtrsim 20\degr{}$ at the location of the planet \citep{Zhu_ea_2015, Bae_ea_2018a, Bae_ea_2018b}. In contrast, spirals driven by buoyancy resonances are much more tightly wound, with pitch angles monotonically decreasing with orbital radius and no significant changes around the location of the planet \citep{Bae_ea_2021}. A second distinction between these two mechanisms is the location at which the perturbations peak. For the case of Lindblad spirals, the most significant deviations from a smooth background are found at a radial distance of a few scale heights from the planet, while for spirals driven by buoyancy resonances, the strength of the deviations is peaking at the location of the planet \citep{Zhu_ea_2015, Bae_ea_2018a, Bae_ea_2018b, Bae_ea_2021}.

If we assume the spirals to be predominantly Lindblad spirals, then the embedded planet needs to be located inside the observed spirals in order to account for the radial trend in $\beta$. This places the planet within the mm~continuum, and thus the two gaps at 73 and 141~au prove to be likely locations. Alternatively, if the spirals are dominated by buoyancy resonances, then the launching radius cannot be constrained by the radial dependence of $\beta$, rather on the strength of the perturbations. As the spirals are only tentatively detected, it is hard to isolate the most significant perturbation, however it seems that the inner edge of the spirals, $\sim\! 220$~au ($1\farcs36$), display the largest azimuthal deviations after the subtraction of an azimuthally averaged background model.

We additionally observe two tightly wound spirals in the gas velocity structure at radii of 162 and 243~au ($1\farcs0$ and $1\farcs5$), at a position angle that connects them to the large variations seen in the gas temperature. These two features have extremely small pitch angles, $\beta \lesssim 5\degr$, and constitute the largest perturbations in the gas velocity. It is therefore more likely that these are associated with buoyancy resonances than Lindblad spirals. In this scenario, the planet responsible for driving these feature would be in the immediate vicinity, $\lesssim 0.2\arcsec$, of the perturbations.


\paragraph{Changes in Rotation Velocity}

In addition to opening a gap in the distribution of large grains, a planet will perturb the local gas pressure profile, primarily by decreasing the total gas surface density, but also through changes in the local thermal structure \citep[e.g.,][]{Teague_ea_2017, Rab_ea_2020}. Pressure perturbations are detectable in the $v_{\phi}$ profile, with pressure minima characterized by a slowing of the rotation inside of the pressure minimum, and a hastening of the rotation outside \citep{Kanagawa_ea_2015, Teague_ea_2018a}. Using  Fig.~\ref{fig:velocity_profiles}d as a guide, we see that pressure minima are qualitatively consistent with the deviations in the $v_{\phi}$ profile at 143, 187, 245 and 339~au ($0\farcs88$, $1\farcs15$, $1\farcs52$ and $2\farcs10$, respectively). We note that changes in emission height can also drive changes in $v_{\phi}$ (see Eqn.~\ref{eq:v_phi}), however \citealt[MAPS~IV]{Law_ea_2020_surf} found no significant sharp changes in the emission height of $^{12}$CO (2-1) in MWC~480, aside from a depression at 66~au ($0\farcs41$). \citet[][MAPS~IV]{Law_ea_2020_surf} reports a tentative wave-like structure in the emission surface between 100 and 180~au, however the location of the proposed peaks and troughs in the emission surface do not align with the features reported here.


\paragraph{Substructure in Molecular Emission}

At these locations of pressure minima, it is unclear whether either the integrated emission \citep[as traced by zeroth moment maps;][MAPS~III]{Law_ea_2020_rad}, or the local gas temperature (probed by the brightness temperature for optically thick lines) will increase or decrease. Extensive thermo-chemical modeling has shown that these changes are highly dependent on both the magnitude of the perturbation in the gas surface density, but also the level of depletion of small grains in the gap which are the dominant coolant for the gas, and any shadowing effects from puffed-up rims of the gap \citep[e.g.,][]{Teague_ea_2017, Facchini_ea_2018, vanderMarel_ea_2018, Alarcon_ea_2020, Rab_ea_2020}. In general, however, there is a growing consensus that for shallow perturbations in the gas surface density, for example those opened by a Saturn mass planet, the local gas temperature will increase due to the drop in opacity towards UV photons, locally enhancing the CO abundance and thus boosting the CO emission \citep{vanderMarel_ea_2018, Alarcon_ea_2020}. Conversely, for more significant gas gaps, for example driven by a Jupiter mass planet, the drop in total gas column will dominated all other effects and the CO emission will drop, due to combination of lower CO abundances and the CO emission tracing a cooler region, closer to the disk midplane \citep{Rab_ea_2020}. Readers are referred to MAPS V \citep{Zhang_ea_2020} for a thorough analysis of the radial profiles of CO column densities in these sources.

As such, we remain agnostic about the expected change in temperature and instead correlate the locations of pressure minima with the local maxima or minima in the $T_B$ profile (Fig.~\ref{fig:Tb_profiles}d). From these locations, it is only the 245~au ($1\farcs52$) location that closely matches a pressure minimum and coincides with a extreme of the $T_B$ profile -- here a minimum. Furthermore, this location is the only location that broadly matches a feature in the integrated intensity, D246, a dip at 246~au ($1\farcs55$). As both of these features are a depression, this would suggest that a larger mass planet, $\sim\!1~M_{\rm Jup}$, would be needed to drive such a perturbation.

\vspace{0.2cm}
Although there is no definitive location for a planet, it appears that the observed features are all broadly consistent with being launched by a planet at $\approx 245$~au ($1\farcs52$). Such a planet would open a gap in both the dust continuum, tentatively seen in Fig.~\ref{fig:MWC480_continuum}, and in the gas, consistent with the pressure minimum interpretation of the $v_{\phi}$ variations, and drop in both the $^{12}$CO brightness temperature profile and integrated intensity at $\approx 245$~au (the integrated intensity map is presented in \citealt[MAPS~III,]{Law_ea_2020_rad} and shown to be azimuthally symmetric, with radial variations on the order a few percent relative to the background value are found, similar to the peak brightness profile). Based on the location and strength of the spiral perturbations, it is likely that a planet would be at a position angle between $270\degr$ and $0\degr$, i.e., in the north-west quarter of the sky.

While there is no clear ``kink'' KPS detected in the CO isotopologue emission at this location (see Fig.~\ref{fig:app:MWC480_12CO_channelmaps} for example), the channel maps are dominated by ``spurs'', features which \citet{Bae_ea_2021} argued were spirals from buoyancy resonances, which may hide a KPS. It is unclear why these spur features dominate in MWC~480, and could be due either to the lower inclination of the disk relative to that of HD~163296, thus increasing the sensitivity of our observations to the vertical motions, or a steeper vertical temperature gradient, more favorable for the generation of these resonances, as suggested by the thermal structured extracted from the data and described in MAPS~IV \citep{Law_ea_2020_surf}. High angular resolution observations able to spatially resolve these features may facilitate the identification of a KPS within these spurs around the location of the proposed planet.

\section{Summary}
\label{sec:summary}

In this work, we have analyzed high spatial and spectral resolution observations of CO isotopologue emission from the disks of HD~163296 and MWC~480. Both disks exhibit a stunning variety of gaseous substructures, traced by local variations in the gas temperature and velocity structure, which we summarize below.

A previously detected `kink' KPS in the $^{12}$CO channel maps of HD~163296 was shown to be much more azimuthally extended than originally thought, with velocity deviations observed to span at least half the disk in azimuth. These features tentatively extend the full azimuth of the disk, connecting to a spiral in the southern side of the disk. Unfortunately, deviations due to the large pressure gradient at the outer edge of the disk prevent a more definitive statement.

At smaller radii in the disk of HD~163296, substantial velocity deviations on the order of $\sim\! 200~{\rm m\,s^{-1}}$ were observed in both $^{12}$CO and $^{13}$CO emission. With the magnitude of the perturbations peaking at a radius of 145~au, consistent with a gap in the continuum emission, a planetary origin is an enticing prospect. The coincidence in the azimuthal angle and that of the aforementioned planet at 260~au, however, may indicate that these deviations are related and will require follow-up observations to disentangle.

In addition to HD~163296, the gas disk of MWC~480 displays an exceptional level of substructure.\ $^{12}$CO, $^{13}$CO and C$^{18}$O emission were found to to show concentric radial variations in their $T_B$ profiles of $\approx \!3$~K, with the radial size of the features growing with radius. On top of this complex background, subtle spiral perturbations were detected in the $^{12}$CO emission at the level of $\approx \!3$~K. Linear fits to the spirals suggested small pitch angles (between 2$\degr$ and 6$\degr$), broadly consistent with predictions for spirals driven by buoyancy resonances. Two large arc-shaped perturbations in the gas velocity structure were found to connect to the spirals arms traced by $T_B$, in addition to a significant vertical velocity perturbation centered at 245~au. A planet at $\approx 245$~au could be responsible for driving the observed perturbations, in addition to a tentative gap at 245~au in the dust continuum.

These observations represent some of the deepest and most well resolved images of CO emission from a protoplanetary disk to date, and highlight the utility of deep, spectral line observations of protoplanetary disks. We would see significant benefits in an improved characterization of these features from future follow-up observations better suited to spatially and spectrally resolve them. These sources are also ideal targets for the upcoming James Webb Space Telescope, as the predicted planets lie at large separations from the central star, where contamination from the stellar PSF is minimized.

\acknowledgements
This paper makes use of the following ALMA data: ADS/JAO.ALMA\#2018.1.01055.L and 2016.1.00484.L. ALMA is a partnership of ESO (representing its member states), NSF (USA) and NINS (Japan), together with NRC (Canada), MOST and ASIAA (Taiwan), and KASI (Re- public of Korea), in cooperation with the Republic of Chile. The Joint ALMA Observatory is operated by ESO, AUI/NRAO and NAOJ. The National Radio Astronomy Observatory is a facility of the National Science Foundation operated under cooperative agreement by Associated Universities, Inc.

V.V.G. acknowledges support from FONDECYT Iniciaci\'on 11180904 and ANID project Basal AFB-170002. A.S.B. acknowledges the studentship funded by the Science and Technology Facilities Council of the United Kingdom (STFC). K.Z. acknowledges the support of the Office of the Vice Chancellor for Research and Graduate Education at the University of Wisconsin – Madison with funding from the Wisconsin Alumni Research Foundation. K.Z., K.R.S., J.H., J.B., J.B.B., and I.C. acknowledge the support of NASA through Hubble Fellowship grant HST-HF2-51401.001, HST-HF2-51419.001, HST-HF2-51460.001-A, HST-HF2-51427.001-A, HST-HF2-51429.001-A, and HST-HF2-51405.001-A awarded by the Space Telescope Science Institute, which is operated by the Association of Universities for Research in Astronomy, Inc., for NASA, under contract NAS5-26555. C.J.L. acknowledges funding from the National Science Foundation Graduate Research Fellowship under Grant DGE1745303. A.D.B. and E.A.B. acknowledges support from NSF AAG Grant \#1907653. K.I.\"O acknowledges support from the Simons Foundation (SCOL \#321183) and an NSF AAG Grant (\#1907653). R.L.G. acknowledges support from a CNES fellowship grant. S.M.A. and J.H. acknowledge funding support from the National Aeronautics and Space Administration under Grant No. 17-XRP17 2-0012 issued through the Exoplanets Research Program. J.D.I. acknowledges support from the Science and Technology Facilities Council of the United Kingdom (STFC) under ST/T000287/1. C.W. acknowledges financial support from the University of Leeds, STFC and UKRI (grant numbers ST/R000549/1, ST/T000287/1, MR/T040726/1). R.T. and F.L. acknowledges support from the Smithsonian Institution as a Submillimeter Array (SMA) Fellow. Y.A. and G.C. acknowledges support by NAOJ ALMA Scientific Research Grant Code 2019-13B and Grant-in-Aid for Scientific Research 18H05222 and 20H05847. J.K.C. acknowledges support from the National Science Foundation Graduate Research Fellowship under Grant No. DGE 1256260 and the National Aeronautics and Space Administration FINESST grant, under Grant no. 80NSSC19K1534. A.S. acknowledges support from ANID/CONICYT Programa de Astronom\'ia Fondo ALMA-CONICYT 2018 31180052. F.M. acknowledges support from ANR of France under contract ANR-16-CE31-0013 (Planet-Forming-Disks) and ANR-15-IDEX-02 (through CDP ``Origins of Life"). Y.B. acknowledges funding from ANR (Agence Nationale de la Recherche) of France under contract number ANR-16-CE31-0013 (Planet-Forming-Disks). H.N. acknowledges support by NAOJ ALMA Scientific Research Grant Numbers 2018-10B and Grant-in-Aid for Scientific Research 18H05441. Y.Y. is supported by IGPEES, WINGS Program, the University of Tokyo. L.M.P.\ acknowledges support from ANID project Basal AFB-170002 and from ANID FONDECYT Iniciaci\'on project \#11181068. T.T. is supported by JSPS KAKENHI Grant Numbers JP17K14244 and JP20K04017.

\facilities{ALMA}

\software{
    \vspace{0.1cm}\\
    \vspace{0.1cm}\texttt{eddy} \citep{eddy}\\
    \vspace{0.1cm}\texttt{emcee} \citep{emcee}\\
    \vspace{0.1cm}\texttt{GoFish} \citep{GoFish}\\
    \vspace{0.1cm}\texttt{matplotlib} \citep{matplotlib}\\
    \texttt{scipy} \citep{scipy}.
    }
    
\bibliography{main}
\bibliographystyle{aasjournal}


\appendix

\section{Creation of Rotation Maps}
\label{sec:app:creation_of_rotation_maps}

Prior to collapsing the data into a rotation map, it is often useful to apply a smoothing kernel along the spectral axis to remove noisy peaks. This can remove high frequency noise and allow for a more accurate determination of the desired statistic, however also can result in subtle deviations being washed out. The Python package \texttt{bettermoments} \citep{Teague_Foreman-Mackey_2018} offers two ways to smooth the data prior to collapsing the cube: convolution with a top hat kernel of a user-defined size, or the use of Savitzky–Golay filter with a user-defined window size and polynomial order. The Savitzky–Golay filter is frequently used in spectroscopy as it retains much of the \emph{shape} of the profile, which can sometimes be lost with other forms of convolution.

In addition to spectrally smoothing the data when making the $v_0$ map, a convolution in the spatial plane after subtracting the model velocity map will remove some noise, aiding in the identification of coherent structures. One major source of structure in these residuals is the feathering due to the channelization of the data.  

\begin{figure}
    \centering
    \includegraphics[width=0.85\textwidth]{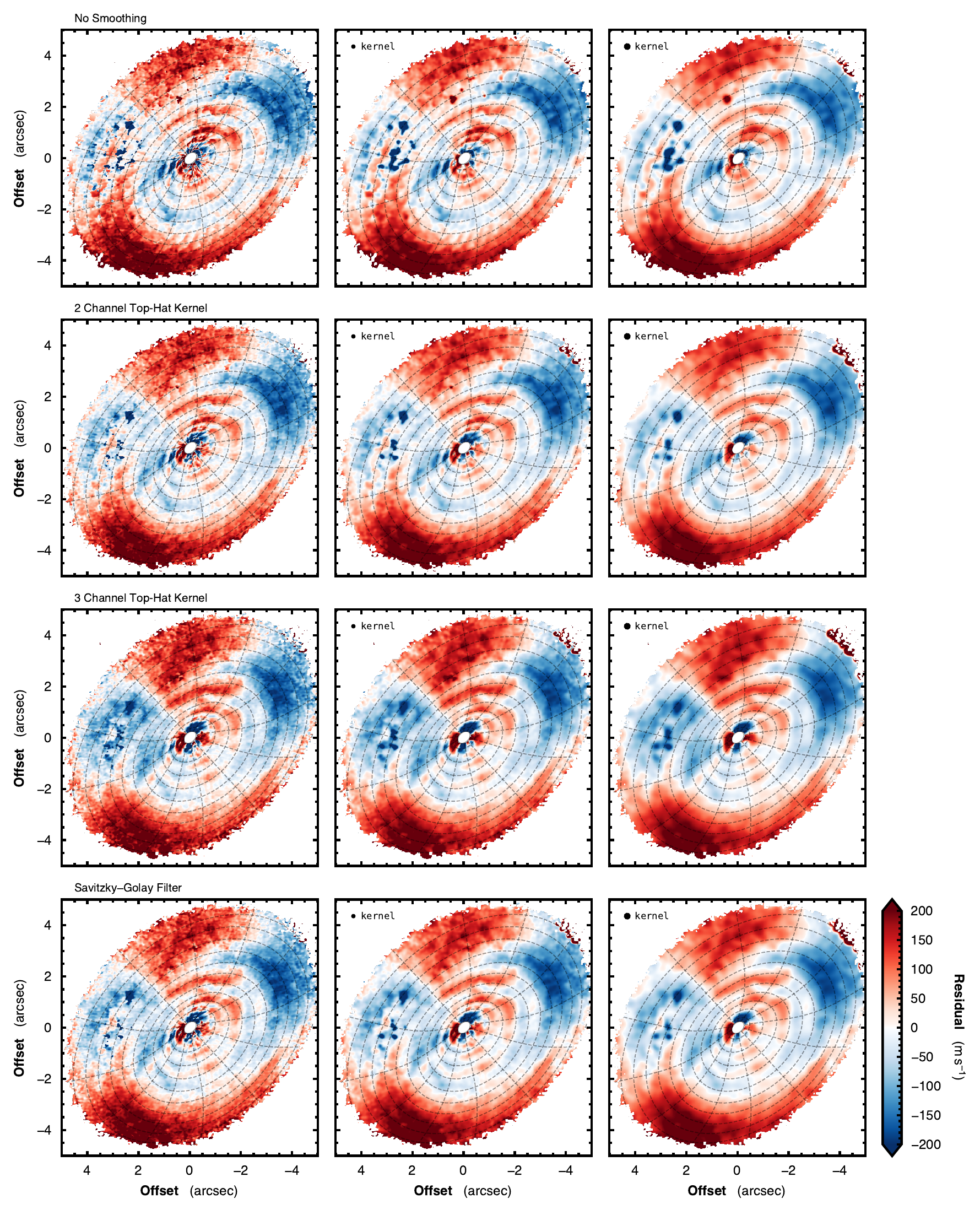}
    \caption{Comparison of the HD~163296 $^{12}$CO $v_0$ map residuals after subtracting a Keplerian rotation model when using different levels of spectral and spatial smoothing. Each row represents a different spectral smoothing: no smoothing, top row; top hat convolutions, second and third rows; Savitzky-Golay filter, bottom row. Each column has a different level of spatial smoothing applied: no smoothing, left column; convolution with a Gaussian kernel with a standard deviation of 1 pixel and 2 pixels, middle and right columns, respectively. The kernel size is plotted in the top left of the appropriate panels.}
    \label{fig:app:HD163296_v0maps_smoothing}
\end{figure}

In the main text, we chose to use no smoothing prior to the collapse of the data as it was found to wash out some of the more subtle deviations. Fig.~\ref{fig:app:HD163296_v0maps_smoothing} demonstrates the impact of the smoothing on the $v_0$ map of $^{12}$CO in HD~163296. The top left panel shows no spectral smoothing, the top right and bottom left panels show convolution with a top-hat kernel with a window size of 2 and 3 channels, respectively, while the bottom right panel shows the Savitzky–Golay filter with a window size of 5 and a 2nd order polynomial. All residual maps have been additionally smoothed with a 2D Gaussian convolution in the spatial plane to further suppress noise. This figure shows that the smoothing does not introduce any new features, however can highlight some of the more prominent features (e.g., the large red arc at a $\sim\!2\arcsec$ offset directly North of the disk center), while simultaneously washing out some of the more subtle detail (e.g., the azimuthal extension both leading and following the aforementioned arc).

\section{The Inclination of MWC~480}
\label{sec:app:MWC480_inclination}

\begin{figure}
    \centering
    \includegraphics{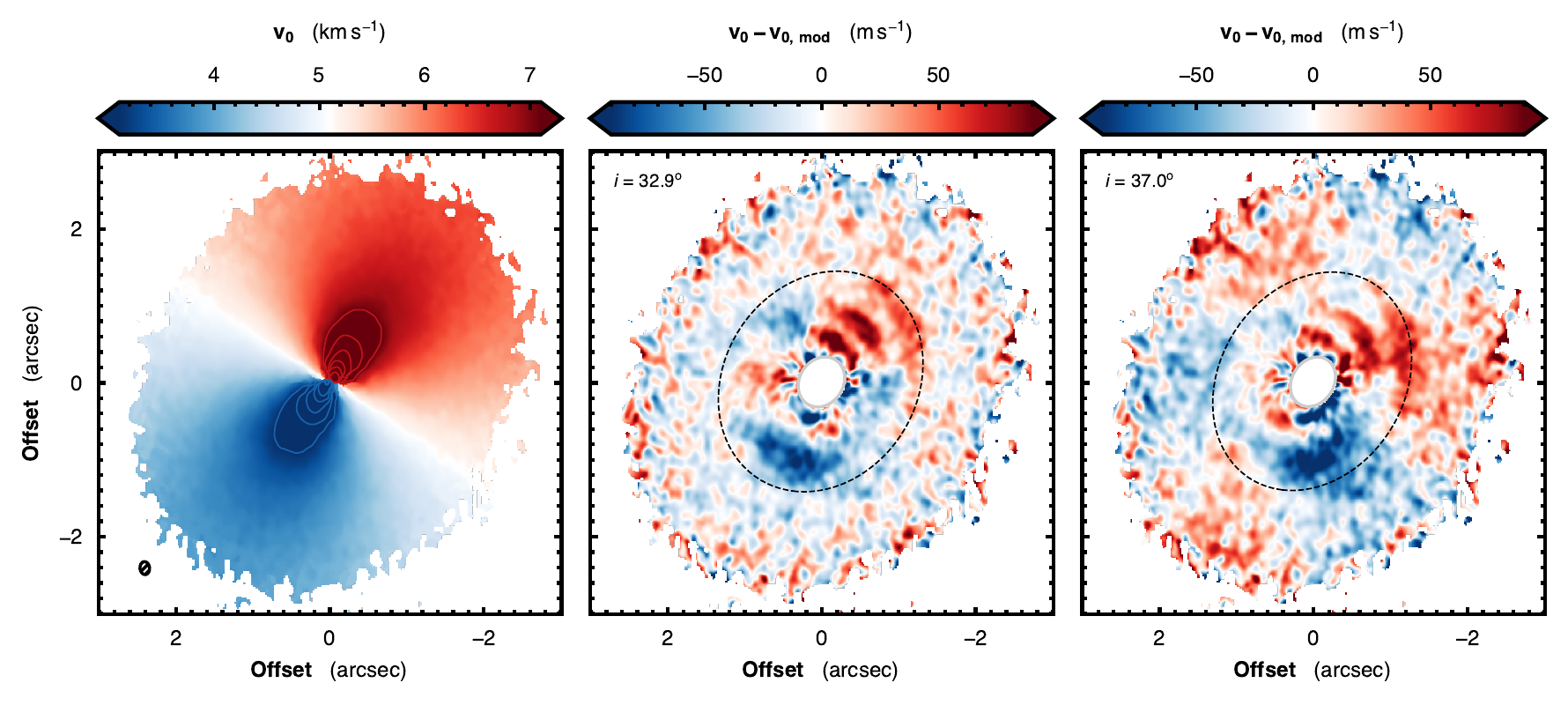}
    \caption{Left: The $v_0$ map of $^{13}$CO (2-1) in MWC~480. The synthesized beam is shown by the ellipse in the lower left corner of the panel. Center: Residuals from a Keplerian model where the inclination was allowed to vary. The fit only considered emission outside a (projected) radius of $1\farcs8$, shown by the dashed black line. Right: Residuals from a Keplerian model where the inclination was fixed to $37\degr$.}
    \label{fig:app:MWC480_13COv0_comparison}
\end{figure}

Within the MAPS collaboration, the inclination of MWC~480 was taken to be $i = 37\degr$, based on the continuum fitting presented in \citet{Liu_ea_2019}. However, when fitting the $v_0$ maps for MWC~480, as described in Appendix~\ref{sec:app:rotation_map_fitting}, it was noticed that this fixed inclination resulted in a rotationally symmetric residual in the outer disk for all three CO isotopologues considered, as shown in the right panel of Fig.~\ref{fig:app:MWC480_13COv0_comparison}. Such residuals are indicative of a mis-specified inclination.

To test this hypothesis, the same $v_0$ fitting procedure was performed for all three CO isotopologues, however letting the inclination to vary. Fitting the entire disk yielding a larger inclination of $\approx 35\degr$, however this still resulted in systematic residuals seen in the outer disk. As the inclination was likely biased by the strong non-Keplerian features in the inner disk (see Section~\ref{sec:rotation_maps:MWC480}), this process was repeated, but fitting only the outer regions of the disk where there were no clear substructures, as shown in Fig.~\ref{fig:MWC480_v0maps}. This resulted in fitting outside $2\farcs5$ for $^{12}$CO, outside $1\farcs8$ for $^{13}$CO and outside $1\farcs5$ for C$^{18}$O. All three fits yielded an inclination of $i \approx 33$, while also removing the systematic residuals in the outer disk, as shown by the central panel in Fig.~\ref{fig:app:MWC480_13COv0_comparison}.

As the emission surfaces described in \citet[MAPS~IV]{Law_ea_2020_surf} were based on an inclination of $37\degr$, we repeat the fitting using the same method with an inclination of $i = 22.4\degr$, with the results of the tapered power law fit shown in Table~\ref{tab:app:v_kep_models}. We find little difference between the emission surfaces reported in \citet[MAPS~IV]{Law_ea_2020_surf}, however find a slightly elevated $^{12}$CO emission surface due to the change in projection from the inclination. For this paper, we adopt these emission surfaces. For a final check, the $v_0$ map fits are performed again, fitting only the outer edge of the disk, however using the updated emission surfaces. No significant change is found for the inclination from $\approx 33\degr$ as the outer edge of the disk is close to being geometrically thin.

The continuum features discussed in \citet[MAPS~III]{Law_ea_2020_rad} were also checked to make sure the updated inclination did not significantly change their locations. Features were found to vary only by a few au as the change in projected radius is small compared to the observed size of the feature.

\section{Rotation Map Fitting}
\label{sec:app:rotation_map_fitting}

In this Appendix, we describe the rotation map fitting procedure with \texttt{eddy} and provide the best-fit results. \texttt{eddy} \citep{eddy} is a wrapper for \texttt{emcee} \citep{emcee}, a Python ensemble sampler for affine-invariant MCMC. In the modeling, five free parameters were considered, the source offset from the phase center, the disk position angle, the host star mass and the systemic velocity: $\{\delta x_0,\, \delta y_0,\, {\rm PA},\, M_*,\, v_{\rm LSR}\}$, while the disk inclination, $i$, and emission surface profiles, $z(r)$, were held constant. For each run, the inner $\approx 0\farcs3$ was masked as beam dilution limited the precision that $v_0$ could be calculated. The outer fitting radius was dictated by the contamination from the rear side of the disk, with the adopted values provided in Table~\ref{tab:app:v_kep_models}. The uncertainty maps produced by \texttt{bettermoments} were adopted as the uncertainties.

To explore the posterior distributions of the free parameters, 32 walkers were used. The walkers took 2,000 steps to burn in (it was found that walkers only needed $\approx 500$ steps to converge), then an additional 2,000 steps to sample the posterior distribution. The posterior distributions for all parameters were found to be Gaussian, with little to no covariance between other parameters. The 16th to 84th percentile ranges were therefore used to estimate the statistical uncertainty on each model parameter, which are provided in Table~\ref{tab:app:v_kep_models}. The $v_0$ maps for the CO isotopologues for HD~163296 and MWC~480 are presented in Figs.~\ref{fig:HD163296_v0maps} and \ref{fig:MWC480_v0maps}, respectively. These include the residuals after subtracting the best Keplerian rotation model. Excellent agreement is found between model parameters for different CO isotopologues. Larger differences found in the dynamical masses can be attributed to the large kinematical deviations found within these disks which will bias the fitting procedure.

We note that this procedure was performed for all sources within the MAPS sample --- IM~Lup, GM~Aur, AS~209, HD~163296, and MWC~480 --- in order to derive dynamical masses for the generation of Keplerian masks (used for imaging the data and extracted spectra). As all other MAPS papers adopted an inclination of $37\degr$ for MWC~480, the published literature value, the fit for MWC~480 was repeated with this larger inclination. This yielded a lower dynamical mass of $M_{*} = 2.1~M_{\sun}$, consistent with the dynamical mass reported in \citet{Simon_ea_2019}.

\begin{deluxetable}{@{\extracolsep{6pt}}rlcccccc@{}}
\tablecaption{Best Fit $v_{\rm kep}$ Models from MAPS Data\label{tab:app:v_kep_models}}
\tabletypesize{\footnotesize}
\tablehead{\\[-8pt]
\nocolhead{} & \nocolhead{} & \multicolumn3c{HD~163296} & \multicolumn3c{MWC~480} \\[3pt]\cline{3-5} \cline{6-8}
\multicolumn2c{Model Parameter} & \colhead{$^{12}$CO \,(2-1)} & \colhead{$^{13}$CO \,(2-1)} & \colhead{C$^{18}$O \,(2-1)} & \colhead{$^{12}$CO \,(2-1)} & \colhead{$^{13}$CO \,(2-1)} & \colhead{C$^{18}$O \,(2-1)}}
\startdata
$\delta x_0$ & (mas)                    & $-20 \pm 1$ & $-16 \pm 1$ & $-11 \pm 1$ & $-19 \pm 1$ & $-13 \pm 1$ & $-9 \pm 1$ \\
$\delta y_0$ & (mas                     & $9 \pm 1$ & $10 \pm 1$ & $16 \pm 1$ & $5 \pm 1$ & $6 \pm 1$ & $10 \pm 1$ \\
$i$ & (\degr{})                         & [46.7] & [46.7] & [46.7] & [$-$32.4] & [$-$32.4] & [$-$32.4] \\
${\rm PA}$ & (\degr{})                  & $312.7 \pm 0.1$ & $312.7 \pm 0.1$ & $312.6 \pm 0.1$ & $327.9 \pm 0.1 $ & $327.5 \pm 0.1$ & $327.5 \pm 0.1$ \\
$M_*$ & ($M_{\odot}$)                   & $2.01 \pm 0.01$ & $1.93 \pm 0.1$ & $1.92 \pm 0.1$ & $2.61 \pm 0.01$ & $2.57 \pm 0.01$ & $2.60 \pm 0.01$ \\
$v_{\rm LSR}$ & (${\rm km\,s^{-1}}$)    & $5.76 \pm 0.01$ & $5.77 \pm 0.01$ & $5.75 \pm 0.01$ & $5.10 \pm 0.01$ & $5.10 \pm 0.01$ & $5.09 \pm 0.01$ \\
$z_0$ & (\arcsec{})                     & [0.39] & [0.12] & [0.17] & [0.30] & [0.11] & [0.07] \\
$\phi$ & (-)                            & [1.85] & [1.50] & [2.96] & [1.32] & [1.17] & [1.18] \\
$r_{\rm taper}$ & (\arcsec{})           & [2.36] & [3.16] & [1.04] & [3.27] & [1.43] & [1.09] \\
$q_{\rm taper}$ & (-)                   & [1.18] & [5.00] & [4.99] & [2.73] & [4.11] & [3.11] \\
$d$ & (pc)                              & [101.0] & [101.0] & [101.0] & [161.8] & [161.8] & [161.8] \\
$r_{\rm fit,\,in}$ & (\arcsec{})        & [0.28] & [0.29] & [0.29] & [0.33] & [0.34] & [0.34] \\
$r_{\rm fit,\,out}$ & (\arcsec{})       & [3.75] & [2.25] & [2.25] & [4.25] & [3.00] & [3.00] \\[3pt]
\enddata
\tablecomments{Uncertainties represent the 16th to 84th percentiles of the posterior distribution. Values in braces were held fixed during the fitting.}
\end{deluxetable}

\section{Channel Maps}
\label{sec:app:channel_maps}

\begin{figure*}
    \centering
    \includegraphics[width=0.85\textwidth]{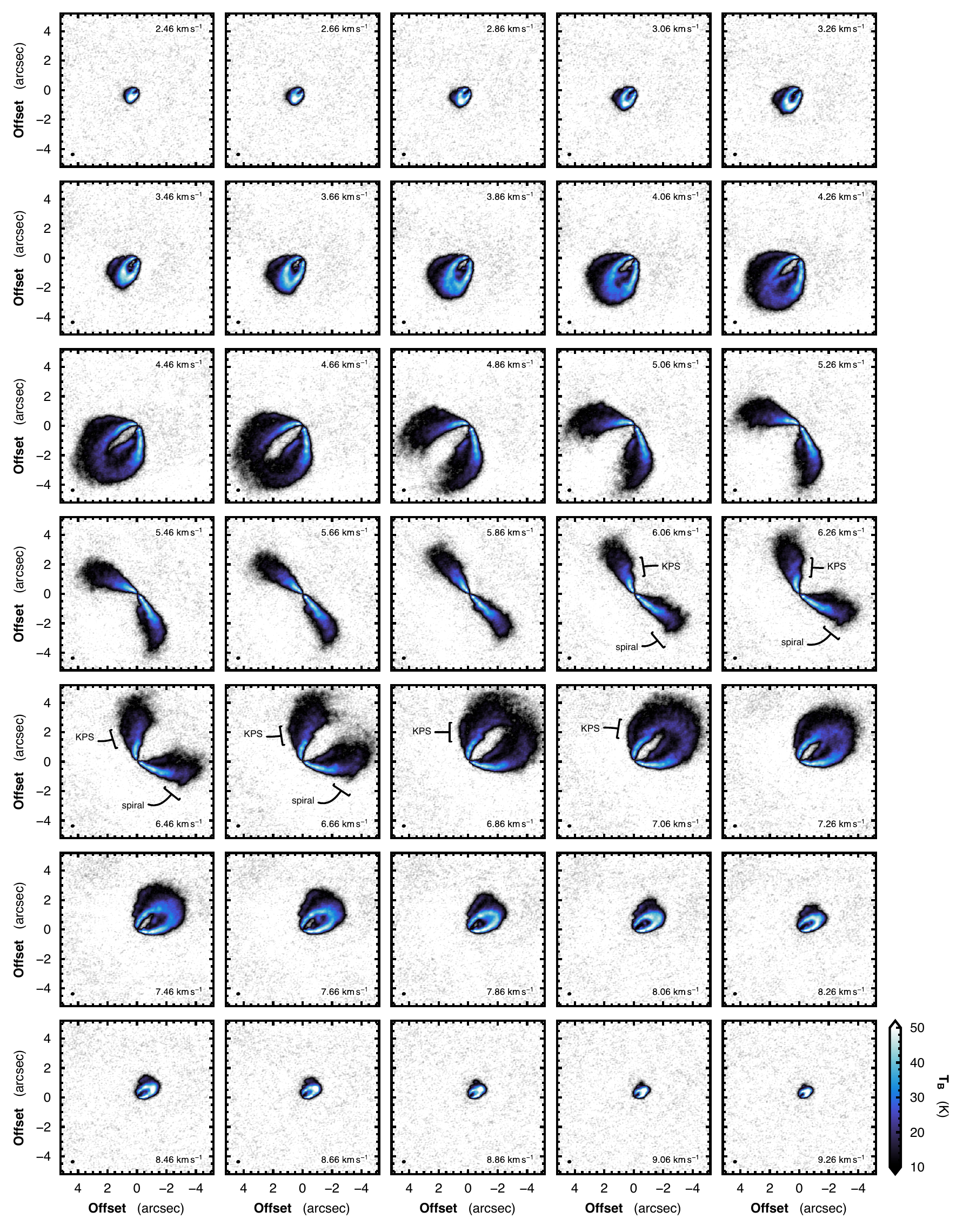}
    \caption{Channel maps of $^{12}$CO $J = 2-1$ emission in HD~163296. The color scaling has been chosen to highlight the morphology of the emission from the line wings. The velocity is given for each channel, as well as the synthesized beam shown in the bottom left of each panel. Regions where the KPS and spiral are most visible are annotated.}
    \label{fig:app:HD163296_12CO_channelmaps}
\end{figure*}

\begin{figure*}
    \centering
    \includegraphics[width=0.85\textwidth]{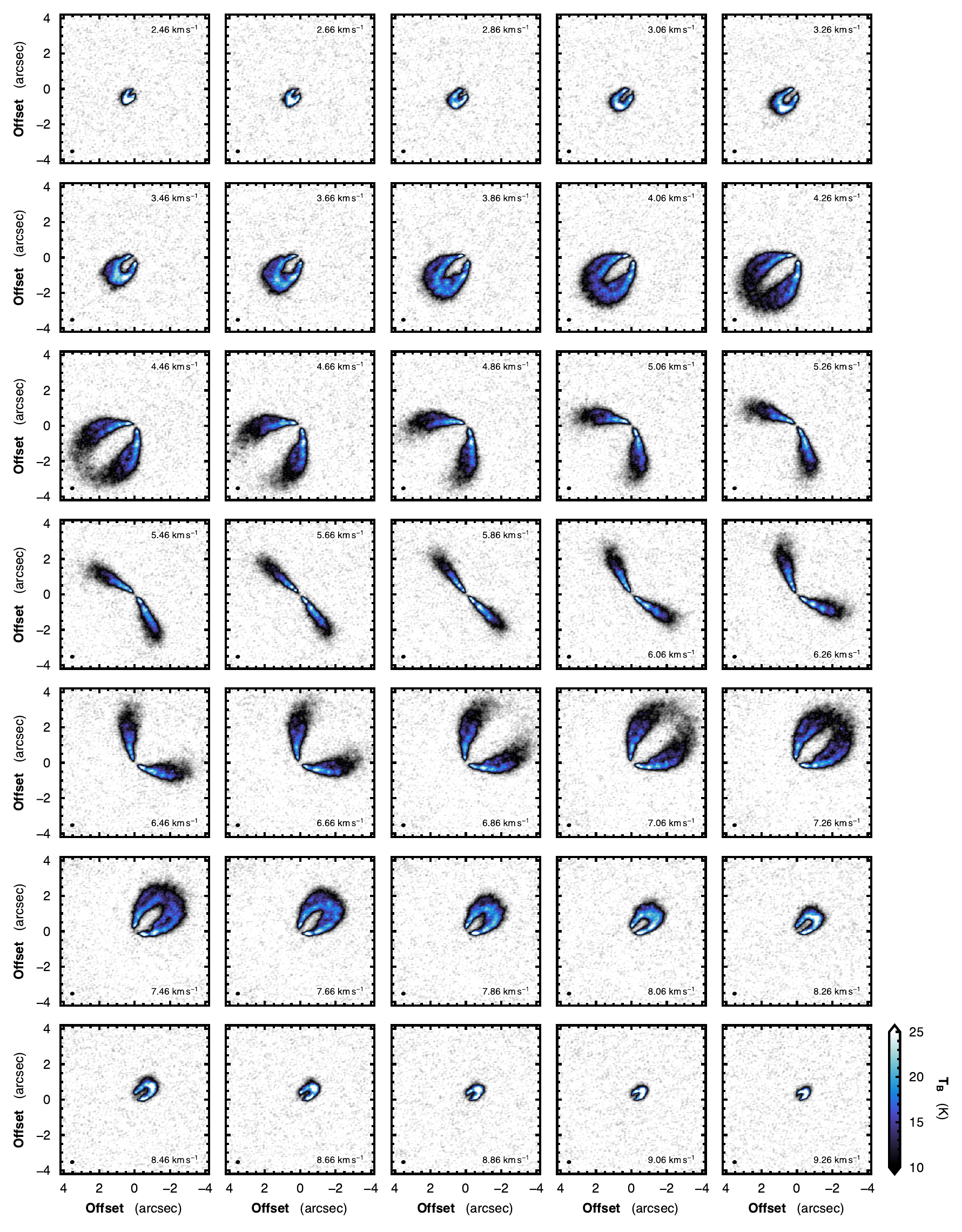}
    \caption{As Fig.~\ref{fig:app:HD163296_12CO_channelmaps}, but for $^{13}$CO $J = 2-1$ emission.}
    \label{fig:app:HD163296_13CO_channelmaps}
\end{figure*}

\begin{figure*}
    \centering
    \includegraphics[width=0.85\textwidth]{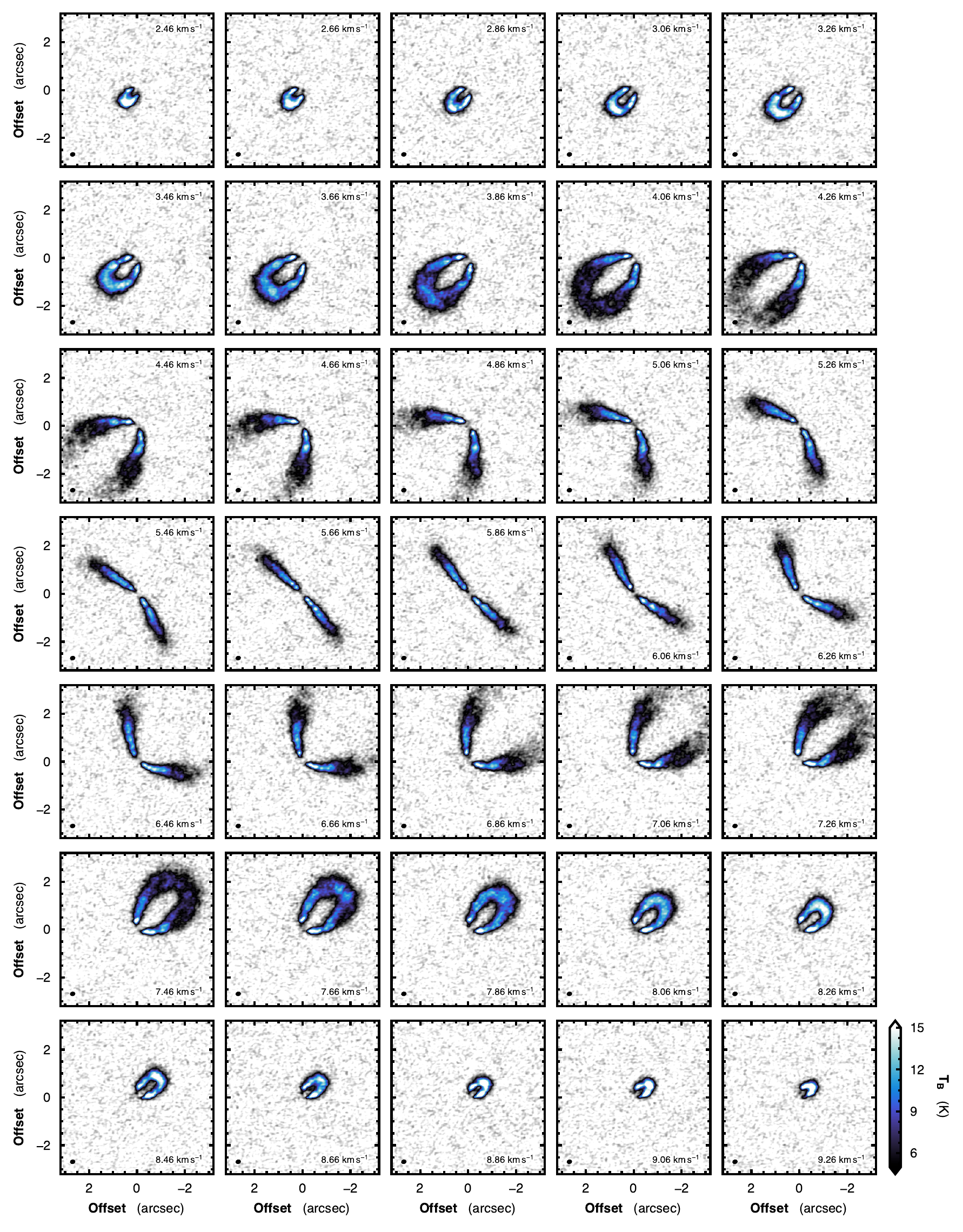}
    \caption{As Fig.~\ref{fig:app:HD163296_12CO_channelmaps}, but for C$^{18}$O $J = 2-1$ emission.}
    \label{fig:app:HD163296_C18O_channelmaps}
\end{figure*}

\begin{figure*}
    \centering
    \includegraphics[width=0.85\textwidth]{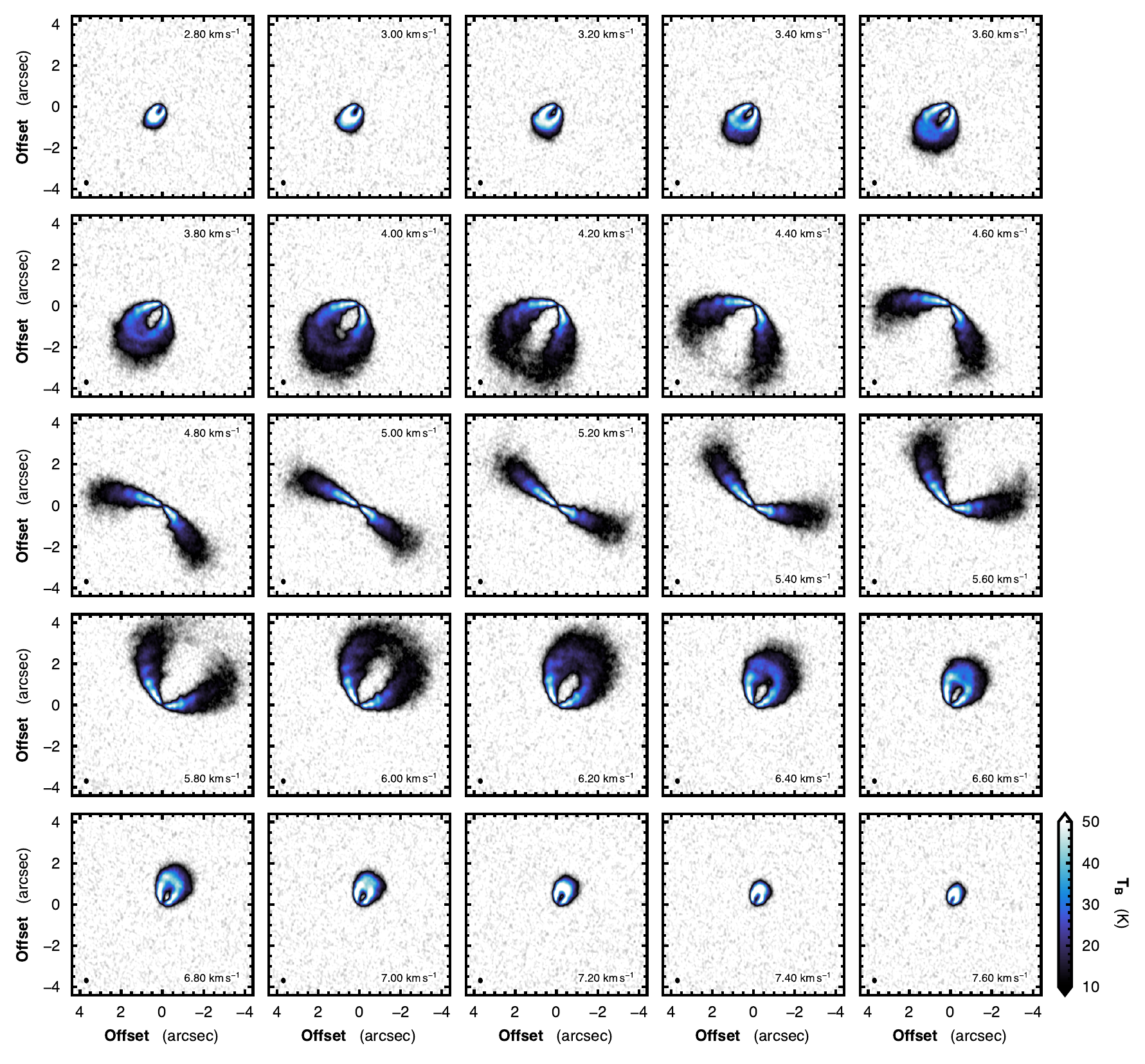}
    \caption{Channel maps of $^{12}$CO $J = 2-1$ emission in MWC~480. The color scaling has been chosen to highlight the morphology of the emission from the line wings. The velocity is given for each channel, as well as the synthesized beam shown in the bottom left of each panel.}
    \label{fig:app:MWC480_12CO_channelmaps}
\end{figure*}

\begin{figure*}
    \centering
    \includegraphics[width=0.85\textwidth]{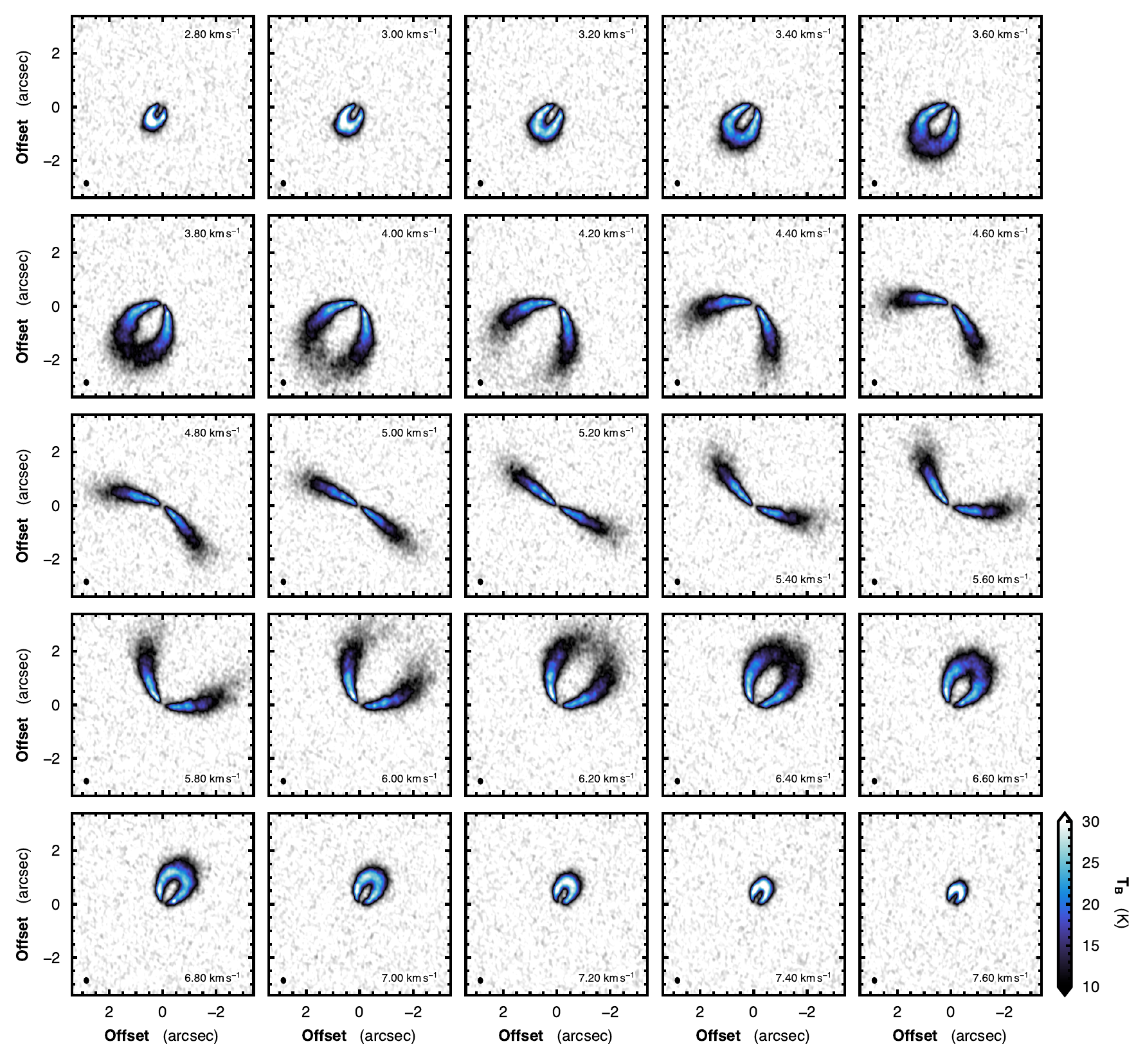}
    \caption{As Fig.~\ref{fig:app:MWC480_12CO_channelmaps} but for $^{13}$CO $J = 2-1$ emission.}
    \label{fig:app:MWC480_13CO_channelmaps}
\end{figure*}

\begin{figure*}
    \centering
    \includegraphics[width=0.85\textwidth]{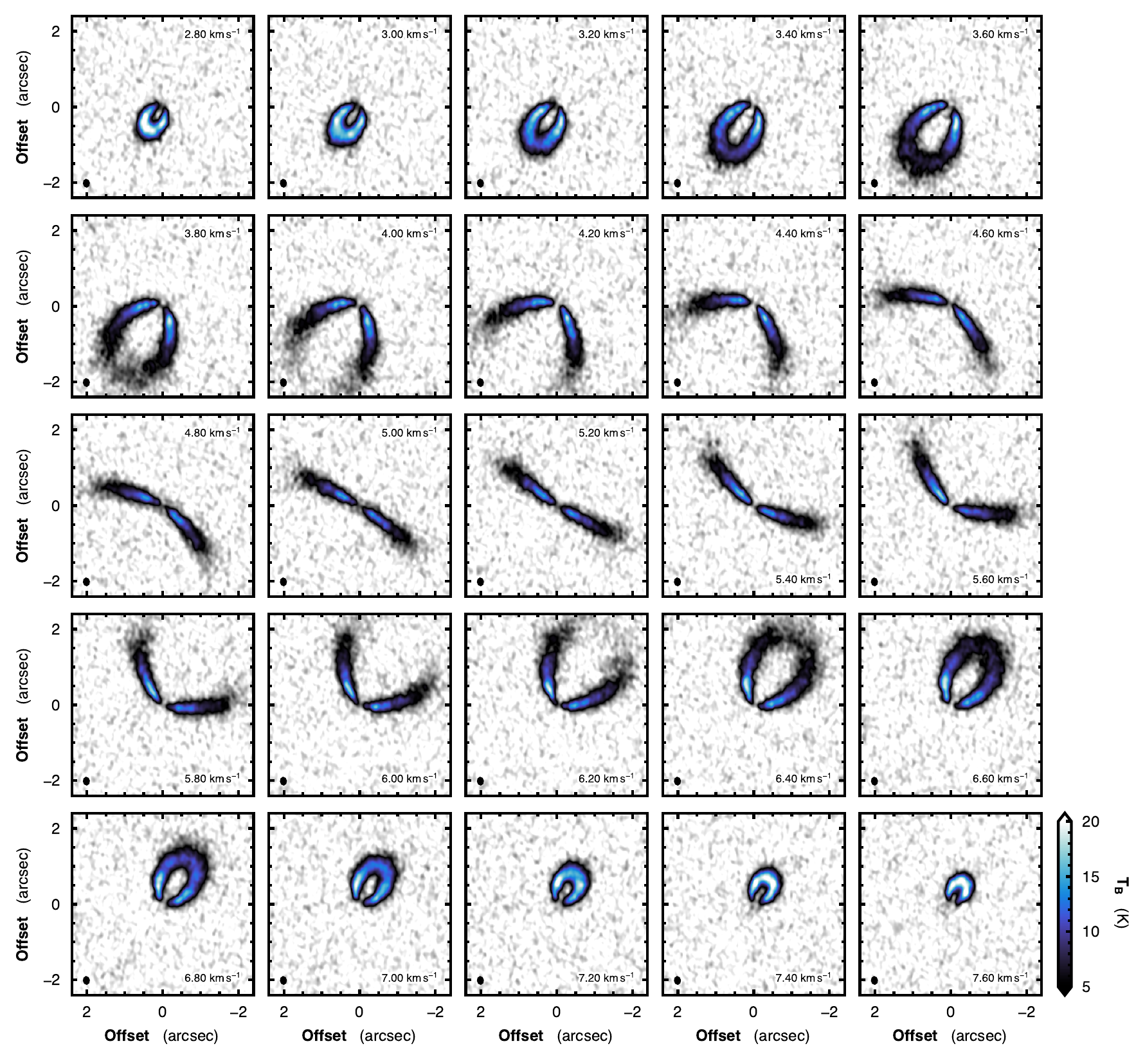}
    \caption{As Fig.~\ref{fig:app:MWC480_12CO_channelmaps} but for C$^{18}$O $J = 2-1$ emission.}
    \label{fig:app:MWC480_C18O_channelmaps}
\end{figure*}

Searching for coherent structures in the residuals when subtracting a rotation model from the line center maps is both quick, and relatively straight forward to compare to maps of the dust continuum. As demonstrated in \citet{DiskDynamics_ea_2020}, the same information is found in the channel maps,\footnote{Indeed, the proposed planet in the disk of HD~163296 was originally identified in the channel maps rather than in the rotation map residuals \citep{Pinte_ea_2018b}.} although presenting a harder challenge for interpretation. The reward to identifying the structures in a channel-map basis is that there is considerably more information compared to a rotation map, which has been first collapsed along the spectral axis. While this paper focused on the collapsed rotation maps, we provide the channel maps in this Appendix. Figs.~\ref{fig:app:HD163296_12CO_channelmaps}, \ref{fig:app:HD163296_13CO_channelmaps} and \ref{fig:app:HD163296_C18O_channelmaps} show the (2-1) transitions of $^{12}$CO, $^{13}$CO and C$^{18}$O in HD~163296, respectively. Figs.~\ref{fig:app:MWC480_12CO_channelmaps}, \ref{fig:app:MWC480_13CO_channelmaps} and \ref{fig:app:MWC480_C18O_channelmaps} show the same molecular lines but for MWC~480.

\end{document}